\documentclass[12pt,journal,draftcls,a4paper,onecolumn]{IEEEtran}

\usepackage[nomarkers,nofiglist,notablist]{endfloat} 

\usepackage{dsfont}
\usepackage{amsmath}
\usepackage{amssymb}
\usepackage{amsfonts}
\usepackage{graphicx}
\usepackage{amsmath}
\usepackage[all,poly]{xy}
\usepackage{multirow}
\usepackage{algorithm}
\usepackage{algorithmic}
\usepackage{color}
\usepackage[noadjust]{cite}
\usepackage{subfigure}
\usepackage{tikz,pgf}
\usetikzlibrary{arrows,automata}
\usepackage{verbatim}
\usetikzlibrary{positioning,shapes.geometric}
\usepackage{url}
\usepackage[T1]{fontenc}

\newcommand{\figwidth}{\columnwidth}

\newcommand{\bzero}{\boldsymbol{0}}

\newcommand{\bmu}{\boldsymbol{\mu}}

\newcommand{\beps}{\boldsymbol{\epsilon}}

\newcommand{\bThe}{\boldsymbol{\Theta}}
\newcommand{\bsig}{\boldsymbol{\sigma}}
\newcommand{\bSig}{\boldsymbol{\Sigma}}
\newcommand{\bgam}{\boldsymbol{\gamma}}
\newcommand{\bGam}{\boldsymbol{\Gamma}}

\newcommand{\bLam}{\boldsymbol{\Lambda}}
\newcommand{\bphi}{\boldsymbol{\phi}}
\newcommand{\bPhi}{\boldsymbol{\Phi}}
\newcommand{\red}{\textcolor[rgb]{1.00,0.00,0.00}}
\newcommand{\blue}{\textcolor[rgb]{0.00,0.00,0.80}}

\def\bsa{{\boldsymbol{a}}}

\def\bsc{{\boldsymbol{c}}}
\def\bsd{{\boldsymbol{d}}}
\def\bse{{\boldsymbol{e}}}

\def\bsk{{\boldsymbol{k}}}

\def\bsm{{\boldsymbol{m}}}

\def\bsp{{\boldsymbol{p}}}

\def\bss{{\boldsymbol{s}}}

\def\bsw{{\boldsymbol{w}}}

\def\bsy{{\boldsymbol{y}}}
\def\bsz{{\boldsymbol{z}}}

\def\bsA{{\boldsymbol{A}}}

\def\bsD{{\boldsymbol{D}}}

\def\bsH{{\boldsymbol{H}}}

\def\bsK{{\boldsymbol{K}}}

\def\bsM{{\boldsymbol{M}}}

\def\bsQ{{\boldsymbol{Q}}}

\def\bsS{{\boldsymbol{S}}}

\def\bsY{{\boldsymbol{Y}}}

\def\dsR{{\mathds{R}}}

\def\calS{{\mathcal{S}}}

\def\calG{{\mathcal{G}}}

\def\calI{{\mathcal{I}}}

\def\calN{{\mathcal{N}}}

\def\calS{{\mathcal{S}}}

\newcounter{algo}
\renewcommand{\thealgo}{\arabic{algo}}

\title{Hyperspectral Unmixing in Presence of Endmember Variability, Nonlinearity or Mismodelling Effects}

\author{Abderrahim Halimi$^{\,1}$\thanks{(1) A. Halimi is with the School of Engineering and Physical Sciences, Heriot-Watt University, Edinburgh U.K.  (e-mail: a.halimi@hw.ac.uk).}, \textit{Member, IEEE,} Paul Honeine$^{\,2}$\thanks{(2) P. Honeine is with the Normandie Univ, UNIROUEN, UNIHAVRE, INSA Rouen, LITIS, Rouen, France  (e-mail: paul.honeine@univ-rouen.fr).}, \textit{Member, IEEE}, Jose Bioucas-Dias$^{\,3}$\thanks{(3) J. M. Bioucas-Dias is with  the  Instituto  de  Telecomunicaç\~oes  and  Instituto  Superior  Técnico,  Universidade  de  Lisboa, Portugal (bioucas@lx.it.pt).}, \textit{Senior Member, IEEE},
\thanks{This work was supported in part by the HYPANEMA ANR Project under Grant ANR-12-BS03-003, and in part by the Portuguese Fundac\~ao para a Ci\^encia e Tecnologia (FCT), grant UID/EEA/5008/2013}}

\begin{document}

\maketitle

\begin{abstract}
This paper presents three hyperspectral mixture models jointly with Bayesian algorithms for supervised hyperspectral unmixing. Based on the residual component analysis model, the proposed general formulation assumes the linear model to be corrupted by an additive term whose expression can be adapted to account for nonlinearities (NL), endmember variability (EV), or mismodelling effects (ME). The NL effect is introduced by considering a polynomial expression that is related to bilinear models. The proposed new formulation of EV accounts for shape and scale endmember changes while enforcing a smooth spectral/spatial variation. The ME formulation takes into account the effect of outliers and copes with some types of EV and NL. The known constraints on the parameter of each observation model  are modeled via suitable priors. The posterior distribution associated with each Bayesian model is optimized using a coordinate descent algorithm which allows the computation of the maximum a posteriori estimator of the unknown model parameters.  The proposed mixture and Bayesian models and their estimation algorithms are validated on both synthetic and real images showing competitive  results regarding the quality of the inferences and the computational complexity when compared to the state-of-the-art  algorithms.
\end{abstract}

\begin{keywords}
Hyperspectral imagery, endmember variability, nonlinear spectral unmixing, robust unmixing, mismodelling effect, Bayesian estimation, coordinate descent algorithm, Gaussian process, Gamma
Markov random field
\end{keywords}

\section{Introduction} \label{sec:Introduction}
Spectral unmixing (SU) of hyperspectral images has been the subject of intensive
interest over the last two decades. This is a source separation problem consisting of recovering the spectral signatures (endmembers) of the materials present in the scene, and quantifying their proportions within each hyperspectral image pixel. The linear mixture model (LMM) is the widely used model for SU mainly because of its simplicity. However, this model can be inappropriate for some hyperspectral scenarios, namely if  there are volumetric scattering, or terrain relief, or  intimate mixtures of materials \cite{Hapke1981}.  Nonlinear mixture models (NLMM) appear then as an alternative to better account for those effects \cite{Heylen2014,Dobigeon2014}. There exists two main families for NLMMs: the first family is signal processing based  and seeks to construct flexible models that can represent a wide range of nonlinearities. The second family is the physical based and includes the intimate mixture models \cite{Hapke1981} and those accounting for multiple scattering such as polynomial \cite{Altmann2012} or bilinear models \cite{Halimi2011TGRS,HalimiIGARSS2011,Nascimento2009,Fan2009,MeganemTGRS2014}. This paper considers physical based nonlinearity while focusing on polynomial/bilinear models.

In both LMM and NLMM, the endmembers are generally assumed fixed in the whole image. This appears as a clear simplification since in many cases, the endmember spectra vary along the image causing what is known as spectral variability or endmember variability (EV) \cite{Somers2011,Zare2014}. Spectral variability has been identified as a relevant source of error in abundance estimation and is attracting growing interest in the hyperspectral community \cite{Somers2011,Zare2014,Halimi_TIP2015}. Many algorithms have been proposed in the literature to describe this variability and they can be gathered into two main classes. The first class considers each physical material as a set or bundle of spectra that are assumed known \cite{Roberts1998,Bateson2000} or estimated from the data \cite{Goenaga2013,Somers2012}. In this class, we find parametric representation of the endmembers such as the multiplication of each endmember by a pixel dependent constant to account for a change of illumination in the observed scene \cite{Nascimento2005b,Veganzones2014,ShawLLJ2003}.
The second class of methods relies on a statistical representation of the endmembers that are assumed to be random vectors with given probability distributions. Two main statistical models of the endmembers have been considered in the literature.  The normal compositional model (NCM) assuming Gaussian distributions for the endmembers \cite{Eches2010ip,Zare2013,Stein2003} and the Beta compositional model \cite{Du_JSTARS2014} that exploits the physically realistic range of the endmember reflectances by assigning them a Beta distribution.  This paper introduces a spectral smooth EV that can be expressed by a Gaussian distributed endmembers resulting in a model that is closely related to NCM.

Mismodelling effects (ME) are also a concern when processing hyperspectral images. These effects can be due to the presence of some physical phenomena such as nonlinearity or endmember variability. However, they can also be due to the propagated errors in the signal processing chain. Indeed, SU is generally performed using three steps: (i) estimating the number of endmembers, (ii) identifying the endmembers using an endmember extraction algorithm (EEA) such as vertex component analysis (VCA) \cite{Nascimento2005}, and N-FINDR \cite{Winter1999} and (iii) estimating the abundances under physical non-negativity and sum-to-one constraints using algorithms such as the fully constrained least squares \cite{Heinz2001}. Many studies consider a supervised unmixing scenario which aims at estimating the abundances while assuming that the two first unmixing steps were successfully  implemented \cite{Heinz2001,BioucasWhispers2010,AltmannTIP2015,ChenTSP2013}. However, an error on the estimated number of endmembers or in their spectra may lead to bad abundance estimates. This problem is considered by some new robust unmixing algorithms that aim at reducing the effect of outliers or mismodelling effects \cite{AltmannTCI2015,FevotteTIP2015}. In this paper, the ME is considered by reducing the effect of smooth spatial/spectral outliers that can be due to the presence of NL, EV or to signal processing chain errors.

The first contribution of this paper is the introduction of a general formulation for the mixture model to account for three phenomena: (i) nonlinearity, (ii) endmember variability or (iii) mismodelling effects. Based on the residual component analysis model \cite{KalaitzisICML2012},  the proposed general formulation assumes the linear model to be corrupted by an additive term whose expression can be adapted to account for the studied phenomenon. This residual term is expressed as a combination of the abundances or the endmembers depending on the studied EV or NL effects. Indeed, the first model studies NL effect which can be modeled by considering a modification to the polynomial term proposed in \cite{AltmannTIP2015}, that depends only on the endmember spectra. The second model considers EV effect by introducing a smooth additive deviation of each endmember from known spectra. Thus, contrary to the NCM described in \cite{Eches2010ip,Zare2013,Halimi_TIP2015}, the proposed model takes into account the smooth spectral and spatial variation of the endmembers in presence of EV. Moreover, and to account for ME,  a third model is introduced for the residual term while accounting for the spatial and spectral smooth properties of the corrupting term. The proposed formulation is therefore general, covers many physical phenomena and can be related to many NL
\cite{Altmann2014,Halimi2011TGRS,Altmann2012,Nascimento2009,Fan2009,MeganemTGRS2014} and EV models \cite{Eches2010ip,Zare2013,Halimi_TIP2015,Veganzones2014}.

The second contribution of this paper is the introduction of three hierarchical Bayesian models associated with each observation model. These hierarchical models introduce prior distributions that enforce  known physical constraints on the estimated parameters such as the sum-to-one and positivity of the abundances, positivity of the nonlinear coefficients and the smooth spectral behavior of EV and ME. Moreover, the spatial correlation of the residual term has been introduced by considering Markov random fields \cite{Rue2005Book,DikmenTASLP2010}. Using the likelihood and the considered prior distributions, the joint posterior distribution of the unknown parameter vector is then derived for each model. The minimum mean square error (MMSE) and maximum a posteriori (MAP) estimators of these parameters cannot be easily computed from the obtained joint posteriors. In this paper, the  MAP estimator is evaluated by considering a coordinate descent algorithm (CDA) \cite{Bertsekas1995,Sigurdsson2014,HalimiTGRS2015} that sequentially  updates the abundances, the noise variances and the residual terms. The proposed Bayesian models and estimation algorithms are validated using synthetic and real hyperspectral images. The obtained results are very promising and show the potential of the proposed mixture and Bayesian models and their associated inference algorithms.

The remainder of this paper in organized as follows. Section \ref{sec:Problem_formulation} introduces the proposed general formulation for the mixture model and its variants to deal with NL, EV and ME effects.
The proposed hierarchical Bayesian models and their estimation algorithms are introduced in Sections \ref{sec:Hierarchical_Bayesian_model} and \ref{sec:Coordinate_descent_algorithm}. Section \ref{sec:Simulation_results_on_synthetic_data} is devoted to testing and  validating the proposed techniques using synthetic images with known ground truth. Section \ref{sec:Simulation_results_on_real_data} shows results obtained using real hyperspectral images. Conclusions and future work are finally reported in Section
\ref{sec:Conclusions}.

\section{Problem formulation}\label{sec:Problem_formulation}
\subsection{Notations}
The variables used in this paper are described as follows:

\begin{tabular}{ll}
$N$        &  number of pixels \\
$R$        &  number of endmembers  \\
$L$        &  number of spectral bands \\
$D$        &  number of the nonlinear coefficients \\
$\bsy_{i,j} \in \dsR^{L \times 1}$   &  pixel located at the $i$th row and $j$th column   \\
$\bsa_{i,j} \in \dsR^{R \times 1}$   &  abundance vector of the pixel $(i,j)$   \\
$\bse_{i,j} \in \dsR^{L \times 1}$   &  noise vector of the pixel $(i,j)$   \\
$\bgam_{i,j} \in \dsR^{D \times 1}$   &  nonlinear coefficients of the pixel $(i,j)$   \\
$\bsc_{i,j} \in \dsR$   &  illumination coefficient of the pixel $(i,j)$  \\
$\bsm_{r} \in \dsR^{L \times 1}$     &  spectrum of the $r$th fixed endmember    \\
%
$\bsc_{i,j} \bsm_{r} \in \dsR^{L \times 1}$     & pixel dependent spectrum of the $r$th endmember   \\
$\bsk_{r,i,j} \in \dsR^{L \times 1}$     &  smooth vector for endmember $r$  \\
$\bsd_{i,j} \in \dsR^{L \times 1}$     &  smooth outlier vector   \\
$\bphi_{i,j} \in \dsR^{L \times 1}$  & residual term of the pixel $(i,j)$ \\
$\bSig \in \dsR^{L \times L}$        & noise covariance matrix \\
$\bsig^2 \in \dsR^{L \times 1}$      & diagonal of the noise covariance matrix \\
$\bsM \in \dsR^{L \times R}$         &  endmember matrix     \\
$\bsS_{i,j} \in \dsR^{L \times R}$   &  endmember matrix of the pixel $(i,j)$   \\
$\bsY \in \dsR^{L \times N}$         &  spectra of the pixels   \\
$\bsA \in \dsR^{R \times N }$        &  abundance matrix  \\
$\beps \in \dsR^{1 \times N }$        &  variance of the residual terms (or their energy) \\
\end{tabular}

\subsection{Mixing model: nonlinearity, endmember variability and mismodelling effects} \label{subsec:Mixture_Models}
The proposed formulation is based on a residual component analysis model \cite{KalaitzisICML2012} that is expressed as the sum of a linear model and a residual term. The general observation model for the $\left( L \times 1\right)$ pixel   spectrum $\bsy_{i,j}$ is given by
\begin{eqnarray}
\bsy_{i,j} & = &  \sum_{r=1}^{R}{a_{r,i,j} \bss_{r,i,j} } + \bphi_{i,j} \left(\bsS_{i,j},\bsa_{i,j}\right) + \bse_{i,j}  \nonumber \\
& = &   \bsS_{i,j}  \bsa_{i,j} +  \bphi_{i,j}\left(\bsS_{i,j},\bsa_{i,j}\right) + \bse_{i,j},
 \label{eqt:GRCA_NL_EV_ME}
\end{eqnarray}
where $\bsa_{i,j}$ is an $\left(R\times1\right)$ vector of abundances associated with the pixel $(i,j)$, $R$ is the number of endmembers,  $\bse_{i,j}   \sim \calN \left(\bzero ,\bSig  \right)$ is an additive centered Gaussian noise with a diagonal covariance matrix $\bSig = \textrm{diag}\left( \boldsymbol{\sigma}^2 \right)$,  $ \boldsymbol{\sigma}^2  = \left(\sigma^2_1,\cdots,\sigma^2_L \right)^T$ is an $\left(L\times 1\right)$ vector containing the noise variances,  $L$ is the number of spectral bands, $\bsS_{i,j}(\bsM)=\bsS_{i,j}$ is the endmember matrix that depends on each pixel to introduce EV effect, $\bsM$ is a fixed endmember matrix that is assumed known (extracted using an EEA) and $ \bphi_{i,j} \left(\bsS_{i,j},\bsa_{i,j}\right)$ is a residual term that might depend on the endmembers or the abundances to  model NL or EV effects. Due  to  physical  constraints,  the  abundance  vector $\bsa_{i,j}= \left(a_{r,i,j},\cdots,a_{R,i,j} \right)^T$ satisfies the following positivity and sum-to-one (PSTO) constraints
\begin{equation}
a_{r,i,j} \geq 0, \forall r \in \left\{1,\ldots,R\right\} \quad
\textrm{and} \quad \sum_{r=1}^{R}{a_{r,i,j}}=1.
\label{eqt:contraints_linear_model}
\end{equation}
Eq. \eqref{eqt:GRCA_NL_EV_ME} shows a general model that can be adapted to account for different physical phenomena. In this paper, we consider three variants dealing with NL, EV or ME. The NL model is designed to deal with the multiple scattering effect that appears in presence of terrain relief, and/or trees. The EV model accounts for the deviation of the endmembers that is commonly observed in presence of vegetation (such as trees or grass), and shadow. It is common to observe the NL and the EV effects simultaneously when analyzing a scene. Therefore, a ME model has been proposed to account for both effects.  
The next sections provide details regarding each of these models.

\subsubsection{Nonlinearity effect} \label{subsec:Nonlinearity_effect}
Nonlinear mixing models provide a useful alternative for overcoming the inherent limitations of the LMM. Indeed, LMM can be inappropriate for some hyperspectral images, such as those containing trees, vegetation or urban areas. Bilinear/polynomial models have shown useful results for these scenes by addressing the problem of double scattering effects. In addition to the LMM terms, these models consider second order interactions between endmembers and neglects the effect of the higher order terms \cite{Halimi2011TGRS,Nascimento2009,Fan2009}. This paper considers the following polynomial/bilinear nonlinear model
\begin{equation}
\bsy_{i,j} =  c_{i,j} \bsM  \bsa_{i,j} +  \bphi_{i,j}^{NL}  \left(\bsM\right) + \bse_{i,j}
 \label{eqt:GRCA_NL}
\end{equation}
where the residual component are similar to \cite{AltmannTIP2015} as follows
\begin{eqnarray}
\bphi_{i,j}^{NL} \left(\bsM\right) & = & c_{i,j}^2 \left( \sum_{k=1}^{R-1}{\sum_{k'=k+1}^{R}{ \gamma_{i,j}^{(k,k')} \sqrt{2} \bsm_{k} \odot \bsm_{k'}   }} \right. \nonumber \\
& + & \left. \sum_{k=1}^{R}{\gamma_{i,j}^{(k)}  \bsm_{k} \odot \bsm_{k} } \right),
 \label{eqt:NL}
\end{eqnarray}
with $\bgam_{i,j}=\left(\gamma_{i,j}^{(1)}, \cdots, \gamma_{i,j}^{(R)}, \gamma_{i,j}^{(1,2)},\cdots,\gamma_{i,j}^{(R-1,R)}\right)^T, \forall i ,j$  is the $\left(D\times1\right)$ vector of positive nonlinear coefficients\footnote{The nonlinear coefficients represent the additional amount of bilinear interactions between the endmembers, thus, they should be positive \cite{Heylen2014,Dobigeon2014}.},
 $D= \frac{R (R+1)}{2}$, $\odot$  denotes the Hadamard (termwise) product,  $\bss_{r,i,j} = c_{i,j} \bsm_{r}, \forall i ,j$, with $c_{i,j}$ a pixel dependent illumination coefficient. 
The model \eqref{eqt:GRCA_NL} generalizes the model \cite{AltmannTIP2015} by including an EV illumination parameter $c_{i,j}$ that accounts for the main spectral variation of endmembers as shown in \cite{Veganzones2014,ShawLLJ2003}. Contrary to the RCA model \cite{Altmann2014}, model \eqref{eqt:GRCA_NL} considers physical positive $\gamma_{i,j}$ (the RCA model \cite{Altmann2014} can be obtained by marginalizing unconstrained $\bgam_{i,j}$ as shown in \cite{AltmannTIP2015}). Note also that \eqref{eqt:GRCA_NL} generalizes the LMM (obtained when  $\bgam_{i,j}=0,$ and $c_{i,j}=1 \forall, i,j$) and has a polynomial-like form as for the bilinear models (GBM \cite{Halimi2011TGRS}, PPNMM \cite{Altmann2012}, Nascimento \cite{Nascimento2009}, Fan \cite{Fan2009} and Meganem \cite{MeganemTGRS2014} models). Note finally that model \eqref{eqt:GRCA_NL} (with no illumination variation) has been studied in \cite{AltmannTIP2015} when considering a Markov chain Monte-Carlo (MCMC) approach and have shown good performance for processing hyperspectral images. However, the MCMC estimation algorithm was computationally expensive, and we consider in this paper a faster algorithm based on a coordinate descent algorithm.

\subsubsection{Endmember variability effect} \label{subsec:Endmember_variability_effect}
\begin{figure}[h!]
\centering
\includegraphics[width=0.9\figwidth]{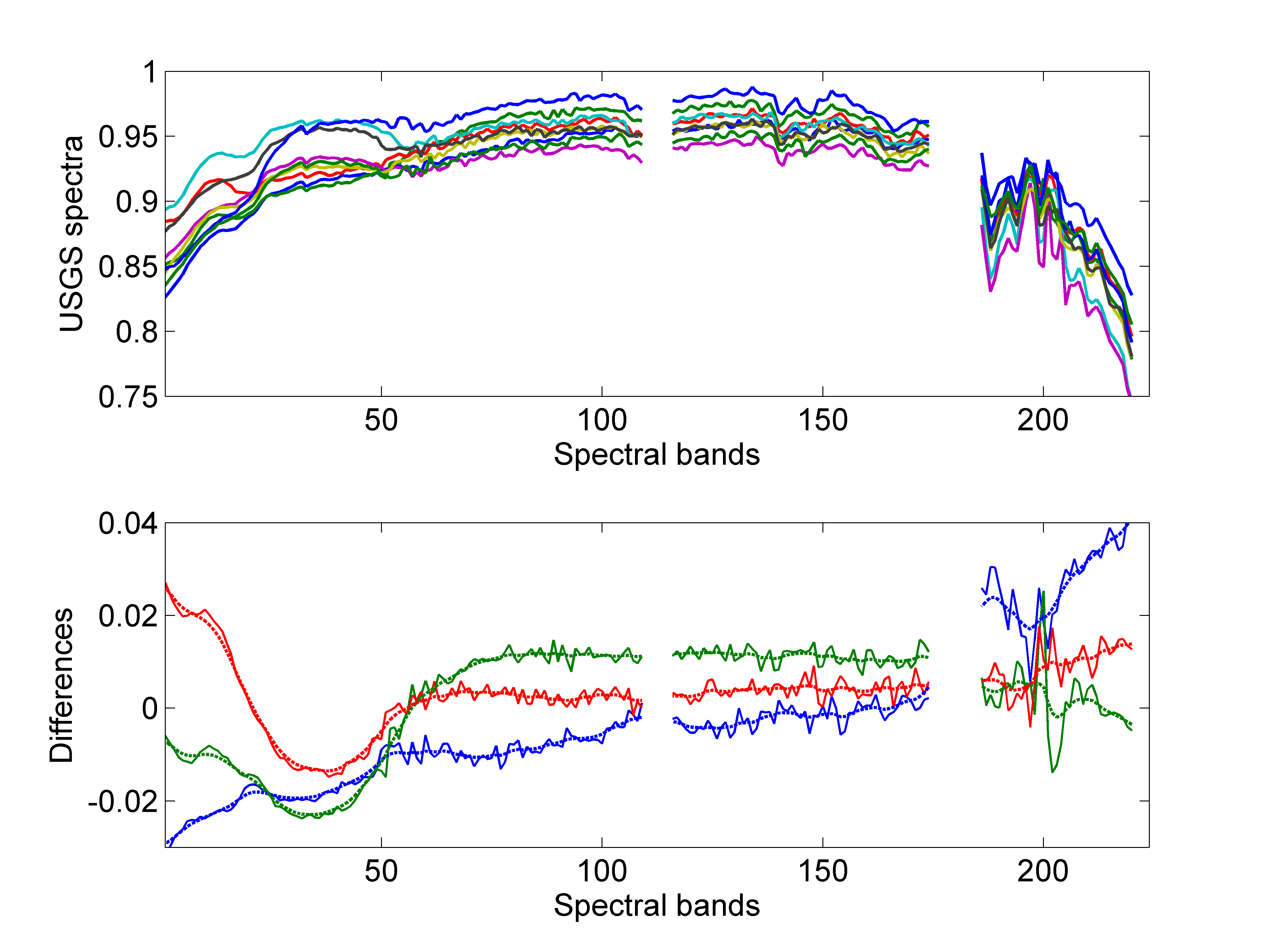}
\caption{(Top) USGS spectra showing endmember variability effect. (Bottom) example of difference between USGS spectra (continuous lines) and their smooth approximation with Gaussian process (dashed lines). } \label{fig:TIP_Justif_Smooth_EV}
\end{figure}

Due to the low spatial resolution of hyperspectral images, the image might represent very large scenes. Therefore, it makes sense to assume that the same material (such as vegetation) might differ with respect to (w.r.t.) the image regions resulting in what it is known as EV.
This variability introduces a modification in the shape and the scale of the endmembers spectrum in each pixel, i.e., $\bss_{i,j}$ depends on the pixel location. As an example, Fig. \ref{fig:TIP_Justif_Smooth_EV} (top) presents spectra associated with the Topaz physical element and extracted from the USGS library. Even though these spectra are associated with the same material, they show some differences which is known as EV effect. To highlight this effect, we compute the average spectrum in each spectral band, and assume that EV is obtained by computing the difference between the spectra of Fig. \ref{fig:TIP_Justif_Smooth_EV} (top) and the average spectrum. An example of the obtained differences is shown in Fig. \ref{fig:TIP_Justif_Smooth_EV} (bottom). This figure clearly shows that the difference between the spectra (which represents the EV effect) can be approximated by smooth functions (see the dashed lines). Therefore, to account for the shape variability of each endmember, each endmember can be approximated by the sum of a fixed spectrum and a smooth spectral function representing EV. This smooth function can be modeled by a parametric approach such as spline, or a statistical approach as Gaussian process (which will be studied in section \ref{subsubsec:Prior_for_the_residual_terms}).
In this paper, we consider the following model for endmember variability
\begin{equation}
 \bss_{r,i,j} =  \bsm_r +  \bsk_{r,i,j},
 \label{eqt:EV1}
\end{equation}
where $\bsk_{r,i,j}$ is a smooth spectral function. This model leads to
\begin{equation}
\bsy_{i,j} =     \bsM  \bsa_{i,j} +  \bphi_{i,j}^{EV} \left(\bsa_{i,j}\right) + \bse_{i,j},
 \label{eqt:GRCA_EV}
\end{equation}
where
\begin{equation}
\bphi_{i,j}^{EV} \left(\bsa_{i,j}\right) = \sum_{r=1}^{R}{a_{r,i,j} \bsk_{r,i,j}}.
 \label{eqt:EV2}
\end{equation}
Note that \eqref{eqt:EV2} does not consider an illumination parameter $c_{i,j}$ since its effect  can be included in the smooth function $\bsk_{r,i,j}$.  Model \eqref{eqt:EV2} relates to state-of-the-art models as follows: (i) it generalizes the LMM that can be retrieved for $\bsk_{r,i,j}=\bzero_L, \forall i,j$, (ii) it generalizes the model \cite{Veganzones2014} by including the effect of shape variability, and (iii) it has a similar formulation as in  \cite{ThouveninTSP2016} while accounting for the spectral smoothness of the residuals.  Note finally that in the special case where $\bsk_{r,i,j}$ is Gaussian distributed (Gaussian process), the model \eqref{eqt:GRCA_EV} will improve the GNCM introduced in \cite{Halimi_TIP2015} by including the smooth behavior of EV (model \eqref{eqt:GRCA_EV} is closely related to the GNCM that can be obtained by marginalizing $\bss_{i,j}$ when considering a Bayesian approach).

\subsubsection{Mismodelling effects (ME) or outliers} \label{subsec:Mismodelling_effect_or_outliers}
The linear model is widely used because of its simplicity. As previously shown, there is a lot of situations where the linear model is not valid because of the presence of variability, nonlinearity or mismodelling effects due to the signal processing chain errors. This section accounts for mismodelling effects or the presence of outliers by considering a residual term that shows spatial and spectral correlations. The observation model is given by
\begin{equation}
\bsy_{i,j} =   c_{i,j} \bsM  \bsa_{i,j} +  \bphi_{i,j}^{ME} + \bse_{i,j}, \;\; \textrm{with} \;\; \bphi_{i,j}^{ME}  =  \bsd_{i,j},
 \label{eqt:GRCA_RL}
\end{equation}
where $\bsd_{i,j}$ is a smooth spectral function. Similarly to the previous models, \eqref{eqt:GRCA_RL} reduces to the LMM when $\bsd_{i,j}=\bzero_L,$ and $c_{i,j}=1, \forall i,j$. Note that other models have been introduced in the literature to account for the effect of outliers such as    \cite{AltmannTCI2015}  that proposed  spatial/spectral  correlated outliers by considering discrete MRF and  \cite{FevotteTIP2015} which proposed a positive sparse outliers that has no spatial or spectral correlation. Note that  \eqref{eqt:GRCA_RL} can be seen as a special case of the EV  model \eqref{eqt:EV2} when  $\bsk_{r,i,j}=\bsk_{r',i,j}, \forall r,r'$ and $c_{i,j}=1$, i.e., the same variability is affecting the different endmembers.  Note also that the NL  model \eqref{eqt:NL} reduces to \eqref{eqt:GRCA_RL} if $\gamma^{(k,k')}= \gamma, \forall k,k'$ since the spectra $\bsm_k \odot \bsm_{k'}, \forall ,k ,k'$ are generally smooth. However, \eqref{eqt:GRCA_RL} is more flexible since it does not consider the positivity constraint (mainly to account for EV).
The next section introduces the proposed hierarchical Bayesian models associated with each observation model.



\section{Hierarchical Bayesian model} \label{sec:Hierarchical_Bayesian_model}
This section introduces a hierarchical Bayesian model for the unknown parameters of models \eqref{eqt:GRCA_NL}, \eqref{eqt:GRCA_EV} and \eqref{eqt:GRCA_RL} which consider different residual terms.
The common unknown parameters of these models include  the ($L \times 1$) diagonal covariance matrix of the noise denoted by $\bSig = \textrm{diag}\left( \bsig^2\right)$, the $(R \times N)$ abundance matrix $\bsA$, the $(N \times 1)$ vector of illumination variability $\bsc$ for the NL \eqref{eqt:GRCA_NL} and ME  \eqref{eqt:GRCA_RL} models, and finally the $(1 \times N)$ variance of the residual terms $\beps$ (or their energy). The remaining parameters depend on the considered residual term as follows:
\begin{itemize}
\item the $(D \times N)$ matrix of nonlinear coefficients $\bGam$ for \eqref{eqt:GRCA_NL}
\item the $(R \times L \times N)$ matrix of endmember variability coefficients $\bsK$ for \eqref{eqt:GRCA_EV}
\item the $(L \times N)$ matrix of mismodelling coefficients $\bsD$ for \eqref{eqt:GRCA_RL}.
\end{itemize}
Table \ref{tab:CDA_Params} presents the parameters (and their dimensions) related to each algorithm. It also shows when a spatial correlation (sc) or a spectral smoothing (ss) is introduced.
\begin{table}[h] \centering
\centering \caption{Estimated parameters of the proposed algorithms. Spatial correlation is denoted by (sc) and  spectral smoothing is denoted by (ss) }
		\begin{tabular}{|c|c|c|c|}
\hline  Parameters  & CDA-NL       & CDA-EV       & CDA-ME \\
\hline  $\bsA$      & $R\times N$  & $R\times N$  & $R\times N$ \\
\hline  $\bGam$     & $D\times N$  &     -        & - \\
\hline  $\bsK$      & -   & $R\times L \times N$ (ss, sc) & - \\
\hline  $\bsD$      & -   &                    -  & $L\times N$ (ss) \\
\hline  $\beps$      & $1\times N$ (sc)  &  -  & $1\times N$ (sc) \\
\hline  $\bsig^2$   & $L\times 1$  & $L\times 1$  & $L\times 1$ \\
\hline  $\bsc$      & $1\times N$  & -            & $1\times N$ \\
\hline
\end{tabular}
\label{tab:CDA_Params}
\end{table}

\subsection{Likelihood} \label{subsec:Likelihood}
Using the observation model \eqref{eqt:GRCA_NL_EV_ME}, the Gaussian properties of the noise sequence $\bse_n$, and exploiting independence between the observations in different spectral bands, yield the following Gaussian distribution for the likelihood
\begin{gather}
\noindent f(\bsy_{i,j}|\bsa_{i,j}, \bSig, \bsS_{i,j}, \bphi_{i,j}) \propto     {\left( \frac{1}{ \prod_{l=1}^{L} \sigma^2_{l} }\right)
}^{\frac{1}{2}} \nonumber \\ \times
 \exp
\left\lbrace -\frac{\left[\bsy_{i,j}- \bsS_{i,j} \bsa_{i,j} - \bphi_{i,j}\right]^T  \bSig^{-1} \left[\bsy_{i,j}-\bsS_{i,j}  \bsa_{i,j} - \bphi_{i,j} \right]}{2}   \right\rbrace. \label{eqt:likelihood}
\end{gather}
The joint likelihood of the observation matrix $\bsY$ is obtained by considering the independence between the observed pixels as follows
\begin{equation}
f(\bsY|\bsA, \bSig, \bsS, \bPhi) \propto
\prod_{i}^{}{\prod_{j}{
f(\bsy_{i,j}|\bsa_{i,j}, \bSig, \bsS_{i,j}, \bphi_{i,j})}}.
\label{eqt:joint_likelihood}
\end{equation}

\subsection{Parameter priors} \label{subsec:Parameter_priors}
This section introduces the prior distributions that we have chosen for the parameters of interest $\bsA$, $\bsc, \bGam, \bsK, \bsD, \bSig$ and $\beps$.

\subsubsection{Abundance matrix $\bsA$} \label{subsubsec:Abundance_matrix}
In order to satisfy the constraints \eqref{eqt:contraints_linear_model}, the abundance vector should live in the following simplex 
\begin{equation}
\calS = \left\lbrace \bsa_{i,j}  \big| a_{r,i,j}\geq 0,
\forall r  \; \textrm{and} \;
\sum_{r =1}^{R}{a_{r,i,j} } = 1 \right\rbrace.
\label{eqt:Simplex_Alpha}
\end{equation}
Since there is no additional information about this parameter vector, we propose to assign a uniform prior in the simplex $\calS$ to the abundances \cite{Dobigeon2008,Halimi2011TGRS}.

\subsubsection{Prior for  $\bsc$} \label{subsubsec:Prior_for_c}
The variation in illumination is introduced by the parameter  $c_{i,j}$ that is pixel-dependent. In many works, this parameter has been fixed to the value $1$, which represents the absence of illumination variability \cite{Halimi2011TGRS,Heinz2001,BioucasWhispers2010}. In this paper, we allow this parameter to fluctuate around the value $1$ to include the effect of illumination. Therefore, we assign the following conjugate Gaussian distribution as a prior for $c_{i,j}$
\begin{equation}
c_{i,j}   \sim \calN  \left(1, \eta^2  \right),
\label{eqt:priorC}
\end{equation}
where  $ \sim$ means ``is distributed according to'', $\eta^2$ is a fixed variance that ensures the value of $c_{i,j}$ to be located around $1$, thus,  avoiding negative values for this parameter in practice ($\eta^2= 0.01$ in the rest of the paper). For simplicity, we denote ``$x|\theta \sim ...$'',  by ``$x \sim ...$'' when the parameter $\theta$ is a user fixed parameter.
Note finally that the joint prior of $\bsc$ is obtained by assuming a priori independence between the coefficients $c_{i,j}$, as follows $ f\left( \bsc  \right) = \prod_{i,j}{f\left( c_{i,j}  \right)}$.

\subsubsection{Prior for the residual terms} \label{subsubsec:Prior_for_the_residual_terms}
\paragraph{Nonlinear coefficients $\bgam_{i,j}$} \label{subsubsec:Nonlinear_coefficients}
Due to physical constraints, the nonlinear coefficients should satisfy the positivity constraint. Similarly to \cite{AltmannTIP2015}, $\bgam_{i,j}$ are assigned the following truncated Gaussian prior
\begin{equation}
\bgam_{i,j} | \epsilon_{i,j}^2 \sim \calN_{{(\dsR+)}^D} \left(\bzero_{D}, \epsilon_{i,j}^2 \mathds{I}_D \right),
\label{eqt:priorGam}
\end{equation}
where $\mathds{I}_D$ denotes the $D\times D$ identity matrix and $\epsilon_{i,j}^2$ is a variance parameter that is pixel dependent. From $\eqref{eqt:priorGam}$, it is clear that this variance is related to the strength of the nonlinearities at the pixel
$(i, j)$ (via the norm  $||\bgam_{i,j}||^2$). Moreover, as in \cite{AltmannTIP2015}, we assume the nonlinear energies to vary smoothly from one pixel to another which will be introduced by considering a specific prior for $\epsilon_{i,j}^2 $, as explained in Section \ref{subsubsec:Abundance_Label_prior}. Note finally that the joint prior of $\bGam$ is obtained by assuming a priori independence between the nonlinear coefficients, as follows $ f\left( \bGam |   \beps\right) = \prod_{i,j}{f\left( \bgam_{i,j} | \epsilon_{i,j}^2\right)}$.

\paragraph{EV coefficients $\bsk_{r,i,j}$} \label{subsubsec:EV_Coefficient}
As previously explained, $\bsk_{r,i,j}$ is a smooth spectral vector. This property can be satisfied by considering a parametric expression for $\bsk_{r,i,j}$ (such as spline), or a statistical approach such as Gaussian process. In this paper, we  assign a Gaussian Markov random field prior \cite{Rue2005Book} for $\bsK_{r}=\{\bsk_{r,i,j}, \forall i,j\}$ that ensures both spectral smoothness and spatial correlation between the residuals. The prior is given by \vspace{-0.2cm}
\begin{eqnarray}
\bsK_{r} & \propto &  \exp{\left[-\sum_{i,j} \left(\frac{ \sum_{(i',j')\in\nu(i,j)} ||\bsk_{r,i,j}-\bsk_{r,i',j'}||^2 }{16 \beta_r^2}  \right.\right.} \nonumber \\
& + & {\left. \left.  
\frac{ \bsk_{r,i,j}^T \bsH^{-1} \bsk_{r,i,j}  }{ 2\alpha_r^2} \right)\right]},  
\label{eqt:priorKjoint}
\end{eqnarray}
where $ ||\cdot|| $ denotes the standard $l_2 $ norm, $\nu(i,j)$ is the neighborhood of the pixel $(i,j)$  (an eight neighborhood structure is considered), and $\bsH$ is an $(L\times L)$ matrix\footnote{When processing real images, some bands are removed because of water absorptions and other atmospheric perturbations.  These bands should also be removed from the covariance matrix by removing the corresponding columns and rows.} representing the squared-exponential covariance function given by  $H(\ell,\ell') = \exp\left[-\frac{\left(\ell-\ell'\right)^2}{\left(L/2\right)^2} \right]$, which introduces the spectral smoothness on $\bsk_{r,i,j}$.
The conditional distribution of $\bsk_{r,i,j}$ given its neighbors is a conjugate Gaussian distribution that can be expressed as follows
\begin{equation}
\bsk_{r,i,j} | \bsk_{r,i',j'}  \sim  \left[\calN \left(\bzero_{L}, \alpha_r^2  \bsH \right) \times \calN \left(\bmu_{r,i,j}, \beta_r^2  \mathds{I}_L \right)\right],
\label{eqt:priorKcond} 
\end{equation}
for $(i',j')\in\nu(i,j)$  and  $\bmu_{r,i,j} = 1/8 \sum_{(i',j')\in\nu(i,j)} \bsk_{r,i',j'}$ is obtained by the average of the neighborhood residuals  of the pixel $(i,j)$.
It is clear from \eqref{eqt:priorKcond} that the first normal (left) introduces the spectral smoothness and the second one (right) introduces the spatial correlations between the residuals. The hyperparameters are fixed by cross-validation, where $\alpha_r^2$ controls the amplitude of the spectral smooth residuals and $\beta_r^2$ the degree of spatial correlation between residuals.  Finally the joint prior of $\bsK$ is obtained by assuming a priori independence between $\bsK_r$, that is $f\left( \bsK \right) = \prod_{r=1}^{R}{f\left( \bsK_r \right)}$.

\paragraph{Mismodelling coefficients $\bsd_{i,j}$} \label{subsubsec:Mismodelling_coefficient}
The spectral vector $\bsd_{i,j}$ should also be smooth. This property is satisfied by considering a Gaussian prior for $\bsd_{i,j}$ ensuring smoothness as follows
\begin{equation}
\bsd_{i,j}|   \epsilon_{i,j}^2 \sim \calN \left(\bzero_{L}, \epsilon_{i,j}^2  \bsH \right),
\label{eqt:priorD}
\end{equation}
where $ \bsH$ is the squared-exponential covariance function described in the previous section. As for the nonlinear coefficients, spatial correlation is introduced by enforcing a smooth variation of the residual energies   $\left(\bsd_{i,j}^T \bsH^{-1} \bsd_{i,j}\right)$. This is achieved by considering a specific prior for $\epsilon_{i,j}^2 $, as explained in Section \ref{subsubsec:Abundance_Label_prior}. Finally the joint prior of $\bsD$ is obtained by assuming a priori independence between the mismodelling coefficients $ f\left( \bsD |   \beps\right) = \prod_{i,j}{f\left( \bsd_{i,j} |  \epsilon_{i,j}^2\right)}$.

\subsubsection{Noise variances} \label{subsubsec:Variance_of_endmembers}
The noise variances are assigned a conjugate inverse gamma distribution given by
\begin{equation}
f\left(\bsig^2  \right) = \prod_{\ell=1}^{L} f\left(\sigma^2_{\ell}  \right), \;\; \textrm{   with    } \sigma^2_{\ell} \sim \calI \calG \left(\varphi_{\ell}, \psi_{\ell}\right), \label{eqt:priorSigma2}
\end{equation}
where $\sigma^2_{\ell}$ are assumed a priori independent. The hyperparameters $\varphi_{\ell}$ and $\psi_{\ell}$ are fixed to approximate the Hysime estimated variances \cite{Bioucas2008}. Indeed, Hysime estimates the noise by subtracting the spectral correlated (smooth) signal from the observed data. This is in agreement with the considered model \eqref{eqt:GRCA_NL_EV_ME} that represents the observations as a sum of a smooth part (linear mixture and residuals) and an independent Gaussian noise. Note finally that the parameters can also be set to $\varphi_{\ell}=\psi_{\ell}=0$ in absence of prior knowledge about $\sigma^2_{\ell}$, leading to a noninformative Jeffreys' prior.


\subsubsection{Energy of the residual term} \label{subsubsec:Abundance_Label_prior}
Due to the spatial organization of hyperspectral images, we expect the energies of the nonlinear coefficients  $\bgam_{i,j}$ and the mismodelling coefficients  $\bsd_{i,j}$ to vary smoothly from one pixel to another. This behavior is obtained by introducing an auxiliary variable $\bsw$ (of size  $N_{\textrm{row}}\times N_{\textrm{col}}$) and  assigning a gamma Markov random field prior (GMRF) for the couple $(\beps,\bsw)$ given by (see \cite{DikmenTASLP2010,AltmannTIP2015} for more details regarding this prior)
\begin{eqnarray}
f\left(\beps,\bsw | \zeta \right)  = &    \frac{1}{Z(\zeta)}  \prod_{(i,j)\in \nu_{\beps}} {\epsilon_{i,j}^{-2(4\zeta+1)}} \nonumber \\
\times & \prod_{(i',j')\in \nu_{\bsw}} {w_{i',j'}^{2(4\zeta-1)}} \nonumber \\
\times & \prod_{((i,j),(i',j'))\in \mathcal{E} } {\exp\left( \frac{-\zeta w_{i',j'}^2}{\epsilon_{i,j}^2} \right)},
\label{eqt:priorEnergies}
\end{eqnarray}
where $Z(\zeta)$ is a normalizing constant,  the partition  $\nu_{\beps}$ (resp. $\nu_{\bsw}$) denotes the collection of variables $\beps$ (resp. $\bsw$), the edge set $\mathcal{E}$ consists of pairs $(i,j)$ representing the connection between the variables and  $\zeta$ is a fixed coupling parameter that controls the amount of spatial smoothness enforced by the GMRF. This prior ensures that each $\epsilon^2_{i,j}$ is connected to four neighbor elements of $\bsw$ and vice-versa.  Note that the energies $\epsilon^2_{i,j}$  are conditionally independent and that a $2$nd order neighbors (i.e., the spatial correlation) is introduced via the auxiliary variables $\bsw$. An interesting property of this joint prior is that the conditional prior distributions of $\beps$ and $\bsw$ reduce to conjugate inverse Gamma ($\calI \calG$) and gamma ($\calG $) distributions as follows \vspace{-0.05cm}
\begin{eqnarray}
\epsilon_{i,j}^2 | \bsw,\zeta \sim   \calI \calG \left(4\zeta, 4\zeta \rho_{1,i,j}(\bsw) \right)
\nonumber \\
w_{i,j}^2 | \beps,\zeta \sim     \calG \left(4\zeta, 1/(4\zeta \rho_{2,i,j}(\beps))     \right),
\label{eqt:CondGam_IGam}
\end{eqnarray}
where \vspace{-0.05cm}
\begin{eqnarray}
\rho_{1,i,j}(\bsw) = (w_{i,j}^2 + w_{i+1,j}^2 + w_{i,j+1}^2 + w_{i+1,j+1}^2)  /4
 \nonumber \\
\rho_{2,i,j}(\beps) = (\epsilon^{-2}_{i,j} + \epsilon^{-2}_{i-1,j} + \epsilon^{-2}_{i,j-1} + \epsilon^{-2}_{i-1,j-1})  /4.
\label{eqt:ParamGam_IGam}
\end{eqnarray}

\subsection{Posterior distributions} \label{subsec:Posterior_distributions}
\begin{figure}
\centering
\subfigure[NL model]{
\begin{tikzpicture}
 nodes %
\node[text centered] (Y) {$\bsY$};
\node[above =0.5 of Y, text centered] (top1) {$ $};
\node[above =0.5 of top1, text centered] (top2) {$ $};
\node[above =0.5 of top2, text centered] (top3) {$ $};

\node[right =1  of top1, text centered] (C) {$\bsc$};
\node[right =0.3  of top1, text centered] (SIG) {$\bsig$};
\node[left =0.3  of top1, text centered] (GAM) {$\bGam$};
\node[left =1  of top1, text centered] (A) {$\bsA$};

\node[draw, rectangle, right  =1.5  of top2, text centered] (ETA) {$\eta$};
\node[left  =2.2  of ETA, text centered] (EPS) {$(\beps,\bsw)$};

\node[draw, rectangle, left   =1.2  of ETA, text centered] (VARPHI) {$\varphi$};
\node[draw, rectangle, right  =0.45 of VARPHI, text centered] (PSI) {$\psi$}; 

\node[draw, rectangle, left =0.25  of top3, text centered] (ZETA) {$\zeta$};
 edges %
\draw[->, line width= 1] (GAM) to  [out=270,in=145, looseness=0.5] (Y);
\draw[->, line width= 1] (SIG) to  [out=270,in=35, looseness=0.5] (Y);
\draw[->, line width= 1] (C) to  [out=270,in=0, looseness=0.75] (Y);
\draw[->, line width= 1] (A) to  [out=270,in=180, looseness=0.75] (Y);

\draw[->, line width= 1] (EPS) to  [out=270,in=145, looseness=0.5] (GAM);

\draw[->, line width= 1] (PSI) to  [out=270,in=35, looseness=0.5] (SIG);
\draw[->, line width= 1] (VARPHI) to  [out=270,in=145, looseness=0.5] (SIG);

\draw[->, line width= 1] (ZETA) to  [out=270,in=35, looseness=0.5] (EPS);

\draw[->, line width= 1] (ETA) to  [out=270,in=35, looseness=0.5] (C);
\end{tikzpicture}}

\subfigure[EV model]{
\begin{tikzpicture}
 nodes %
\node[text centered] (Y) {$\bsY$};
\node[above =0.5 of Y, text centered] (top1) {$ $};
\node[above =0.5 of top1, text centered] (top2) {$ $};
\node[above =0.5 of top2, text centered] (top3) {$ $};

\node[right =0.35  of top1, text centered] (SIG) {$\bsig$};
\node[left =0.25  of top1, text centered] (GAM) {$\bsK$};
\node[left =1.1 of top1, text centered] (A) {$\bsA$};

\node[draw, rectangle, left  =0.8  of top2, text centered] (ALP) {$\alpha_r$};
\node[draw, rectangle, right  =0.35  of ALP, text centered] (Beta) {$\beta_r$};

\node[draw, rectangle, right   =0.0  of top2, text centered] (VARPHI) {$\varphi$};
\node[draw, rectangle, right  =0.45 of VARPHI, text centered] (PSI) {$\psi$}; 

 edges %
\draw[->, line width= 1] (GAM) to  [out=270,in=145, looseness=0.5] (Y);
\draw[->, line width= 1] (SIG) to  [out=270,in=35, looseness=0.5] (Y);
\draw[->, line width= 1] (A) to  [out=270,in=180, looseness=0.75] (Y);

\draw[->, line width= 1] (Beta) to  [out=270,in=45, looseness=0.5] (GAM);
\draw[->, line width= 1] (ALP) to  [out=270,in=145, looseness=0.5] (GAM);

\draw[->, line width= 1] (PSI) to  [out=270,in=35, looseness=0.5] (SIG);
\draw[->, line width= 1] (VARPHI) to  [out=270,in=120, looseness=0.5] (SIG);


\end{tikzpicture}} \hspace{0.4cm}
\subfigure[ME model]{
\begin{tikzpicture}
 nodes %
\node[text centered] (Y) {$\bsY$};
\node[above =0.5 of Y, text centered] (top1) {$ $};
\node[above =0.5 of top1, text centered] (top2) {$ $};
\node[above =0.5 of top2, text centered] (top3) {$ $};

\node[right =0.85  of top1, text centered] (C) {$\bsc$};
\node[right =0.25  of top1, text centered] (SIG) {$\bsig$};
\node[left =0.25  of top1, text centered] (GAM) {$\bsD$};
\node[left =1.05  of top1, text centered] (A) {$\bsA$};

\node[draw, rectangle, right  =1.4  of top2, text centered] (ETA) {$\eta$};
\node[left  =2.1  of ETA, text centered] (EPS) {$(\beps,\bsw)$};
\node[draw, rectangle, left   =1.1  of ETA, text centered] (VARPHI) {$\varphi$};
\node[draw, rectangle, right  =0.45 of VARPHI, text centered] (PSI) {$\psi$}; 

\node[draw, rectangle, left =0.2  of top3, text centered] (ZETA) {$\zeta$};
 edges %
\draw[->, line width= 1] (GAM) to  [out=270,in=145, looseness=0.5] (Y);
\draw[->, line width= 1] (SIG) to  [out=270,in=35, looseness=0.5] (Y);
\draw[->, line width= 1] (C) to  [out=270,in=0, looseness=0.75] (Y);
\draw[->, line width= 1] (A) to  [out=270,in=180, looseness=0.75] (Y);
 
\draw[->, line width= 1] (EPS) to  [out=270,in=145, looseness=0.5] (GAM);

\draw[->, line width= 1] (PSI) to  [out=270,in=35, looseness=0.5] (SIG);
\draw[->, line width= 1] (VARPHI) to  [out=270,in=145, looseness=0.5] (SIG);
\draw[->, line width= 1] (ZETA) to  [out=270,in=35, looseness=0.5] (EPS);

\draw[->, line width= 1] (ETA) to  [out=270,in=35, looseness=0.5] (C);
\end{tikzpicture}}
\caption{DAGs for the parameter and hyperparameter priors (the user fixed parameters appear in boxes). (a) the NL model, (b) the EV model and (c) the ME  model.}
\label{fig:DAGs}
\end{figure}
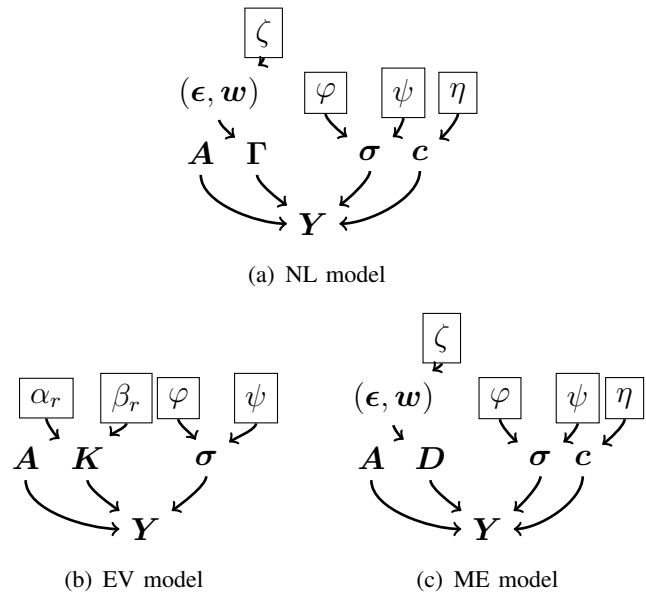

The proposed Bayesian models associated with each observation model are summarized in the directed acyclic graphs (DAGs) displayed in Fig. \ref{fig:DAGs}. The parameters of interest are $\bThe_{\textrm{NL}} = \left(\bsA,\bGam,\bsig,\bsc,\beps,\bsw \right)$,  $\bThe_{\textrm{EV}} = \left(\bsA,\bsK,\bsig \right)$, and  $\bThe_{\textrm{ME}} = \left(\bsA,\bsD,\bsig,\bsc,\beps,\bsw \right)$. The joint posterior distribution of these Bayesian models can be computed from the following Bayes' rule
\begin{equation}
f\left(\bThe | \bsY \right)  \propto f(\bsY|\bThe)  f\left(\bThe  \right),
\label{eqt:Joint_Posterior}
\end{equation}
with
\begin{eqnarray}
f\left(\bThe_{\textrm{NL}} \right) & = &   f\left(\bsA\right) f\left(\bsig\right) f\left(\bsc\right) f\left(\bGam| \beps \right) f\left(\beps,\bsw\right)  \nonumber \\
f\left(\bThe_{\textrm{EV}} \right) & = &   f\left(\bsA\right) f\left(\bsig\right) f\left(\bsK \right)   \nonumber \\
f\left(\bThe_{\textrm{ME}} \right) & = &   f\left(\bsA\right) f\left(\bsig\right) f\left(\bsc\right) f\left(\bsD| \beps \right) f\left(\beps,\bsw\right),
\label{eqt:Priors}
\end{eqnarray}
where $\propto$ means ``proportional to'' and we have assumed a priori independence between the parameters of each model. The MMSE and MAP estimators associated with the posterior \eqref{eqt:Joint_Posterior} are not easy to determine. In this paper, and akin to \cite{HalimiTGRS2015}, we propose to evaluate the MAP estimator by using an optimization technique maximizing the posterior \eqref{eqt:Joint_Posterior} w.r.t. the parameters of interest (or equivalently, minimizing the negative log-posterior $\mathcal{F}(\bThe)= - \textrm{log} [f(\bThe  | \bsY)]$ denoted as ``cost function'' in the rest of this paper).

\section{Coordinate descent algorithms} \label{sec:Coordinate_descent_algorithm}
This section describes the optimization algorithms maximizing the posteriors \eqref{eqt:Joint_Posterior} associated with the NL, EV and ME models w.r.t. the parameters of interest. This provides the MAP estimator of $\bThe_{\textrm{NL}}$,  $\bThe_{\textrm{EV}}$, and  $\bThe_{\textrm{ME}}$.
Because of the large number of parameters to estimate for each model, we propose three coordinate descent algorithms \cite{Bertsekas1995,Sigurdsson2014,HalimiTGRS2015} that sequentially update the different parameters associated with each model, denoted as, CDA-NL, CDA-EV and CDA-ME. For each algorithm and in each step, the posterior distribution is maximized w.r.t. one parameter, the other being fixed. Thus, the algorithm iteratively updates each parameter by maximizing its conditional distribution as follows (see also the Appendix):
\begin{itemize}
\item Conditional of $\bsA$: truncated Gaussian distribution (whose one maximum is obtained with  SUNSAL-FCLS\footnote{SUNSAL-FCLS satisfies the PSTO constraints while SUNSAL-CLS only ensure the positivity constraint.} \cite{BioucasWhispers2010})
\item Parameters of the residual term
\begin{itemize}
\item Conditional of $\bGam$: positive truncated Gaussian distribution (whose one maximum is obtained with SUNSAL-CLS \cite{BioucasWhispers2010})
\item Conditional of $\bsK$: Gaussian distribution (analytical expression of the mean)
\item Conditional of $\bsD$: Gaussian distribution (analytical expression of the mean)
\end{itemize}
\item Conditional of  $\beps$ : inverse gamma distribution (analytical expression of the maximum)
\item Conditional of  $\bsw$ :  gamma distribution (analytical expression of the maximum)
\item Conditional of  $\bsig^2$ : inverse gamma distribution (analytical expression of the maximum)
\item Conditional of  $\bsc^{\textrm{ME}}$: Gaussian distribution (analytical expression of the mean)
\item Conditional of  $\bsc^{\textrm{NL}}$: uncommon distribution (see the Appendix).
\end{itemize}
Regarding the sequence generated by the coordinate descent algorithm, the proposition 2.7.1  in  \cite{Bertsekas1995} asserts that its limit points are stationary points of \eqref{eqt:Joint_Posterior} provided that the maximum of that function w.r.t. $\bThe$ along each coordinate is unique. This is easily checked for all the parameters except for $\bsc^{\textrm{NL}}$. Indeed, the cost function writes as a $4$-order polynomial w.r.t. $\bsc^{\textrm{NL}}$ (leading to $3$ possible maxima) and we have chosen the one that maximizes it in the interval $[0.2, 3]$. Algo. \ref{alg:Gradient_descent_algorithm} gathers the main steps of the proposed three algorithms. If a line is involved in a specific algorithm, it will begin by its name. For example, line $11$ is only executed when considering the algorithms CDA-NL and CDA-ME.  The cost function is not convex, thus, the solution obtained might depend on the initial values that should be chosen carefully. 
Therefore, the abundances $\bsA$ are initialized with SUNSAL-FCLS \cite{BioucasWhispers2010}, the residual terms are initialized by $0$, the noise variance is initialized by  HYSIME \cite{Bioucas2008}, the illumination coefficient $\bsc$ is initialized by considering the sum of the abundances that are estimated using only the positivity constraint with SUNSAL-CLS \cite{BioucasWhispers2010}. With these initializations, the proposed algorithm reached minima of ``good quality'' in the considered simulations (see Sections \ref {sec:Simulation_results_on_synthetic_data} and \ref{sec:Simulation_results_on_real_data}). Therefore, the CDA algorithms constitute a good balance between computational efficiency (obtained by solving the simple problems associated with each descent step) and the quality of the solution (experimentally observed when considering a good initialization).
Finally, we have empirically observed that the algorithm is less sensitive to the initialization when beginning the decent with an abundance update (as in Algo. \ref{alg:Gradient_descent_algorithm}). 
\begin{algorithm}
\caption{Coordinate descent algorithm} \label{alg:Gradient_descent_algorithm}
\begin{algorithmic}[1]
       \STATE Estimate $\bsM$ using an EEA  (VCA \cite{Nascimento2005}, N-FINDR \cite{Winter1999}, ...)
       \STATE \underline{Initialization}
       \STATE Initialize parameters  $\bsA^{(0)}$ (SUNSAL-FCLS), $\bGam^{(0)}=\bzero_{D,N}$,  $\bsK^{(0)}=\bzero_{L,R,N}$, $\bsD^{(0)}=\bzero_{L,N}, \bsc^{(0)}$ (SUNSAL-CLS), $\bSig^{(0)}$ (HYSIME), $\beps^{(0)}$ and $t=1$
       \STATE conv$=0$,
       \STATE \underline{Parameter update}
       \WHILE{conv$=0$}
               \STATE Update $\bsA^{(t)}$ with  SUNSAL-FCLS
               \STATE CDA-NL: Update $\bGam^{(t)}$ with SUNSAL-CLS
			   \STATE CDA-EV: Update $\bsK^{(t)}$ by a standard least squares (analytical expression)
			   \STATE CDA-ME: Update $\bsD^{(t)}$ by a standard least squares (analytical expression)
               \STATE CDA-NL and CDA-ME: Update $(\beps, \bsw)^{(t)}$  (analytical expression)
               \STATE Update $\bSig^{(t)}$  (analytical expression)
               \STATE CDA-ME: Update $\bsc^{(t)}$  by a standard least squares (analytical expression)
               \STATE CDA-NL: Update $\bsc^{(t)}$  by the resolution of a $3$rd order polynomial
               \STATE Set conv$=1$ if the convergence criteria are satisfied
               \STATE t = t + 1
       \ENDWHILE
\end{algorithmic}
\end{algorithm}
\subsubsection{Stopping criteria} \label{subsubsec:Convergence_diagnosis}
Algo. \ref{alg:Gradient_descent_algorithm} is an iterative algorithm that requires the definition of some stopping criteria. In this paper, we have considered four criteria and the algorithm is stopped if one of them is satisfied. The first criterion compares the new value of the cost function to the previous one and stops the algorithm if the relative error between these two values is smaller than a given threshold, i.e.,
\begin{equation}
| \mathcal{F} \left(\bThe^{t+1}\right)-\mathcal{F} \left(\bThe^{t}\right) | \leq  \xi_1 \mathcal{F} \left(\bThe^{t}\right),
\label{eqt:criteria1}
\end{equation}
where $|.|$ denotes the absolute value and the cost function $\mathcal{F} \left(\bThe\right) = - \textrm{log} \left[f\left(\bThe | \bsY \right) \right]$ is the negative log-posterior. The second criterion evaluates
the new abundance values and stops the algorithm if the following condition  is satisfied \cite{Madsen2004}
\begin{eqnarray}
 \left\| \bsA^{(t+1)}-\bsA^{(t)}   \right\|_F \leq  \xi_2 \left( \left\|\bsA^{(t)}\right\|_F  + \xi_2\right).
\label{eqt:criteria1_2}
\end{eqnarray}
where $ \left\|.\right\|_F $ denotes the Frobenius norm.
The third criterion apply the same condition to the residual term ($\bGam$ or $\bsK$ or $\bsD$) while considering a different threshold $\xi_3$.
The last criterion is based on a maximum number of iterations $T_{\textrm{max}}$. These thresholds have been fixed as follows $(\xi_1,\xi_2,\xi_3,T_{\textrm{max}}) = (10^{-5},10^{-6},10^{-6},500)$ for CDA-NL, $(\xi_1,\xi_2,\xi_3,T_{\textrm{max}}) = (5\times 10^{-6},10^{-4},10^{-6},500)$ for CDA-EV   and $(\xi_1,\xi_2,\xi_3,T_{\textrm{max}}) = (10^{-5},10^{-6},10^{-11},500)$ for CDA-ME. The next sections  study the behavior of the proposed algorithms when considering synthetic and real images.

\section{Simulation results on synthetic data} \label{sec:Simulation_results_on_synthetic_data}
This section evaluates the performance of the proposed algorithms with synthetic data. The first part introduces the criteria used for the evaluation of the unmixing quality. In the second part, we evaluate and compare the performance of the proposed algorithms with the state-of-the-art algorithms when considering different unmixing scenarios. In the third part, we analyze the behavior of the proposed algorithms when varying the number of endmembers, spectral bands, pixels and the signal-to-noise ratio (SNR).

\subsection{Evaluation criteria} \label{subsec:Evaluation_criteria}
For synthetic images, the abundances are known and the unmixing quality can be evaluated by using the root mean square error: $\textrm{RMSE}\left(\bsA\right) = \sqrt{\frac{1}{N\, R}\sum_{n=1}^{N}
\left\| \bsa_{n}-\hat{\bsa}_{n} \right\|^{2}}$.   
The unmixing performance can also be evaluated by considering the reconstruction error: $\textrm{RE}= \sqrt{\frac{1}{N\,L}  \sum_{n=1}^{N} \left\| \hat{\bsy}_n-\bsy_n  \right\|^2}$ and the spectral angle mapper $\textrm{SAM}  = \frac{1}{N}  \sum_{n=1}^{N} \arccos \left(
\frac{\hat{\bsy}_n^T \bsy_n}{\left\|\bsy_n\right\| \; \left\|\hat{\bsy}_n  \right\|}\right)$ criteria, where $\arccos(\cdot)$ is the inverse cosine operator and $\bsy_n$, $\hat{\bsy}_n$ denotes  the $\#n$th measured and estimated pixel spectra.

\subsection{Performance of the proposed algorithms} \label{subsec:Performance_of_the_proposed_algorithm}
This section evaluates the performance of the proposed unmixing algorithms when considering different mixture models. Four synthetic images of size  $100 \times 100$ pixels and $L=207$ spectral bands have been generated using 
$R=3$ endmembers corresponding to spectral signatures available in the ENVI software library \cite{ENVImanual2003}. All images have been corrupted by  i.i.d. Gaussian noise (with SNR=$25$ dB) for a fair comparison with SU algorithms using this assumption. The images have been generated using different mixture models as follows
\begin{itemize}
\item Linear model: the image $I_1$ has been generated according to the LMM model with abundances uniformly distributed in the simplex $\calS$. The illumination coefficient increases linearly from the left of the image to its right in the interval $[0.9,1.15]$.
\item Nonlinearity: the image $I_2$ has been generated according to $4$ linear/nonlinear models. For this, an image partition into $4$ classes has been generated by considering a Potts-Markov random field (with granularity parameter $\beta= 0.8$). The four spatial  classes are associated with the LMM, NL model \eqref{eqt:GRCA_NL} (with $\epsilon^2 = 0.1$), GBM (with random nonlinear coefficients in $[0.8,1]$) and PPNMM (with $b=0.5$), respectively. The illumination coefficient increases linearly from the left of the image to its right in the interval $[0.9,1.15]$. Finally, to  make the SU  even more challenging, we have considered a highly mixed scenario by generating the abundance using a Dirichlet distribution whose parameters are selected randomly in the interval $[1,20]$.
\item Endmember variability: the image $I_3$ has been partitioned into $4$ classes (the same classification map as for $I_2$). In each class, we have generated a set of endmembers by considering the model \eqref{eqt:EV1} and $ \bsk_{r,i,j} \sim   \calN \left(\bzero_{L}, \alpha^2  \bsH \right)$, with $\alpha^2 = 0.005, \forall r,i,j$. Therefore, the image will be composed by four sets of endmembers, each one associated with a spatial region. To  make the SU  even more challenging, we have considered a highly mixed scenario by generating the abundance using a Dirichlet distribution whose parameters are selected randomly in the interval $[1,20]$.
\item Mismodelling effects:  the image $I_4$ has been partitioned into $2$ classes (by merging class $1$ with $2$, and $3$ with $4$ of the classification map of $I_2$).  Pixels of the first class have been generated according to the ME model \eqref{eqt:GRCA_RL} with $\epsilon^2=0.002$. The pixels of class $2$ have been generated with $4$ endmembers to simulate the effect of a bad estimation of the number of endmembers that will be fixed to $R=3$ for all unmixing algorithms. The illumination coefficient increases linearly from the left of the image to its right in the interval $[0.9,1.15]$. Finally, to make the SU  even more challenging, we have considered a highly mixed scenario by generating the abundance using a Dirichlet distribution whose parameters are selected randomly in the interval $[1,20]$.
\end{itemize}

These images are processed using different unmixing strategies that are compared to the proposed algorithms. For all algorithms, we have assumed the endmembers to be known and we have considered the ENVI spectra used to create the images. The studied unmixing algorithms are
\begin{itemize}
\item Linear unmixing: the abundances are estimated using the FCLS algorithm \cite{Heinz2001} and the SUNSAL-CLS\footnote{The SUNSAL algorithm is run with the parameters suggested in \cite{BioucasWhispers2010}, i.e., $\lambda=0$, maximum of iterations $= 200$ and tolerance $= 10^{-4}$.} algorithm \cite{BioucasWhispers2010}.
\item Nonlinear unmixing: the abundances are estimated using the MCMC-RCA algorithm \cite{Altmann2014} and the SKHYPE algorithm \cite{ChenTSP2013}
\item Endmember variability: we have considered the  automated endmember bundles (AEB) algorithm proposed in \cite{Somers2012}. In addition to the ENVI spectra, the endmember library includes endmembers extracted using VCA  in  $10\%$ (resp. $20\%$) image subset when processing synthetic (resp. real) images. For each pixel, the $R$ endmembers that provide the smallest RE are selected.
\end{itemize}
The next sections analyse the obtained results w.r.t. each synthetic image.

\subsubsection{Linear model} \label{subsubsec:Linear_model}
Table \ref{tab:Results_Synth_I1_3_4} shows the obtained results when considering the LMM based image $I_1$. Considering the abundance RMSE, and as expected, SUNSAL-CLS provides better results than FCLS because of the variation of the illumination parameter $\bsc$ for this image. Indeed, varying $\bsc$ can be seen as a relaxation of the sum-to-one constraint. This variability also affects RCA-MCMC that presents reduced performance when compared to the other nonlinear based algorithms (CDA-NL and SKhype). The proposed algorithms show very good results especially CDA-NL and CDA-ME which validate their use for linear mixture model. Except FCLS, all algorithms provide good reconstruction spectra which is highlighted by the RE and SAM criteria. This table highlights that the LMM based algorithms are the fastest ones (this result is valid for all images) followed by the CDA-NL algorithm which is $\approx 100$ times faster than the MCMC based algorithm for this image. Overall, CDA algorithms  provide a reduced computational times which shows their efficiency for linear unmixing. Note that CDA-EV requires more computational times mainly because it estimates large matrices $\bsK$. Finally, the obtained results confirm the good behavior of the proposed CDA algorithms when processing LMM based images.

\subsubsection{Nonlinearity} \label{subsubsec:Nonlinearity}
Table \ref{tab:Results_ImNL} shows the obtained results when considering image $I_2$. This table also provides the obtained RMSE associated with each spatial class, i.e., each nonlinear mixture model (the results of class $1$ are similar to those obtained for $I_1$). As expected, the linear based algorithms provide poor results which highlights the need for more elaborated algorithms to deal with the NL effect present in the image. However, it can be seen that removing the sum-to-one constraint in SUNSAL-CLS improves considerably the performance w.r.t. FCLS especially for GBM and PPNMM. This suggest that part of the nonlinearity can be interpreted as a variation of illumination, which is in agreement with \cite{MeganemTGRS2014,Altmann2014} that suggest removing this constraint when considering NL models. The best performance are obtained by considering nonlinear-based algorithms. More precisely, CDA-NL is robust to the different nonlinearities affecting the data and provides the best RMSEs for $I_2$. Note also that CDA-ME shows an intermediate performance between the NL models and the linear/EV  based models. Regarding the computational times, CDA-NL requires less times than SKhype to achieve the unmixing, and again outperforms RCA-MCMC (by a factor of $20$) that shows expensive computational cost. These results confirm the good behavior of CDA-NL when considering NL images. They also show that CDA-ME is an intermediate solution between NL and LMM based algorithms since it requires less computational times than NL algorithms while it shows better results than LMM algorithms.

\subsubsection{Endmember variability} \label{subsubsec:Endmember_variability}
The results associated with $I_3$ are summarized in Table \ref{tab:Results_Synth_I1_3_4}. Again, LMM based algorithms suffer from the presence of EV effect.  As expected, the best performance in terms of abundance RMSE and RE are obtained with the CDA-EV algorithm that is especially designed to deal with the EV effect. Similar good performance are obtained by CDA-ME algorithm which confirms its  robustness with respect to the observed mixture model (LMM, NL or EV). Note that AEB provides bad RMSE results even though it shows a reduced RE. This result shows that a reduced RE does not ensure a good estimation of the abundances. Note finally that apart from the LMM based algorithms, the CDA algorithms show competitive computational times when compared to the remaining algorithms. These results confirm the good performance of the CDA-EV and the robustness of CDA-ME when processing images with EV.

\subsubsection{Mismodelling effects} \label{subsubsec:Mismodelling_effects}
This section analyses the obtained results when processing image $I_4$ that is corrupted by mismodelling effects (presence of outliers and a fourth endmember). For this scenario, CDA-ME provides the best results in term of abundance RMSEs, RE and SAM as shown in Table \ref{tab:Results_Synth_I1_3_4}. Note that CDA-EV also shows interesting results with a reduced computational cost. Nonlinear algorithms suffer from this corruption and show high abundance RMSEs. These results confirm the good performance of CDA-ME in terms of unmixing quality and computational times.

To summarize, the proposed CDA algorithms show reduced computational cost for all images. The best results associated with an image are obtained when considering the corresponding CDA algorithm (for example, CDA-EV is best for EV-based images). CDA-ME is robust to the different mixture models that can affect hyperspectral images.

\begin{table}[h] \centering
\centering \caption{Results on synthetic data (image $I_1$, $I_3$ and $I_4$) .}
\begin{tabular}{|c|c|c|c|c|c|}
\cline{3-6}
\multicolumn{2}{c|}{}                & \multicolumn{3}{c|}{$(\times 10^{-2})$}  &   Time     \\
\cline{3-5} \multicolumn{2}{c|}{}    &   RMSE    &  RE   &  SAM  &  (s)   \\
\hline \multirow{3}{*}{$I_1$}  & FCLS       &  $7.78$         & $3.58$          & $6.24$         & \blue{$1.2$} \\
\cline{2-6}                          & SUNSAL-CLS &  $3.42$         & \blue{$2.27$}   & \blue{$5.62$}  & \red{$0.1$} \\
\cline{2-6}\multirow{2}{*}{(LMM,}    & SKhype     &  $1.41$         & $-$             & $-$            & $541$ \\
\cline{2-6}\multirow{3}{*}{$K=1$)}   & RCA-MCMC   &  $4.31$         & $-$             & $-$            & $6737$ \\
\cline{2-6}                          & AEB        &  $33.8$         & $2.31$          & \blue{$5.62$}  & $1507$ \\
\cline{2-6}                          & CDA-NL     &  \red{$1.34$}   & \blue{$2.27$}   & \blue{$5.62$}  & $45$ \\
\cline{2-6}                          & CDA-EV     &  $3.27$         & \red{$2.24$}    & \red{$5.55$}   & $142$ \\
\cline{2-6}                          & CDA-ME     &  \blue{$1.35$}  & \blue{$2.27$}   & \blue{$5.62$}  & $88$ \\
\hline
\hline \multirow{3}{*}{$I_3$}  & FCLS       &  $10.22$         & $2.91$         & $6.0$          & \blue{$1.3$} \\
\cline{2-6}                          & SUNSAL-CLS &  $10.16$         & $2.53$         & $5.93$         & \red{$0.14$} \\
\cline{2-6}\multirow{2}{*}{(RCA-EV,} & SKhype     &  $7.69$          & $-$            & $-$            & $623$ \\
\cline{2-6}\multirow{3}{*}{$K=4$)}   & RCA-MCMC   &  $13.33$         & $-$            & $-$            & $11025$ \\
\cline{2-6}                          & AEB        &  $19.05$         & \blue{$2.45$}  & $5.74$         & $1465$ \\
\cline{2-6}                          & CDA-NL     &  $10.02$         & $2.53$         & $5.94$         & $520$ \\
\cline{2-6}                          & CDA-EV     &  \red{$3.60$}    & \red{$2.35$}   & \blue{$5.53$}  & $230$ \\
\cline{2-6}                          & CDA-ME     &  \blue{$4.29$}   & \red{$2.35$}   & \red{$5.51$}   & $316$ \\
\hline
\hline \multirow{3}{*}{$I_4$}  & FCLS       &  $12.59$         & $5.01$         & $7.19$         & \blue{$1.21$} \\
\cline{2-6}                          & SUNSAL-CLS &  $12.65$         & $2.91$         & $6.94$         & \red{$0.12$} \\
\cline{2-6}\multirow{2}{*}{(RCA-ME,} & SKhype     &  $11.68$         & $-$            & $-$            & $374$ \\
\cline{2-6}\multirow{3}{*}{$K=4$)}   & RCA-MCMC   &  $18.20$         & $-$            & $-$            & $6029$ \\
\cline{2-6}                          & AEB        &  $32.17$         & $2.88$         & $6.05$         & $1240$ \\
\cline{2-6}                          & CDA-NL     &  $11.85$         & $2.88$         & $6.88$         & $394$ \\
\cline{2-6}                          & CDA-EV     &  \blue{$9.74$}   & \blue{$2.36$}  & \blue{$5.71$}  & $279$ \\
\cline{2-6}                          & CDA-ME     &  \red{$6.07$}    & \red{$2.29$}   & \red{$5.56$}   & $315$ \\
\hline
\end{tabular}
\label{tab:Results_Synth_I1_3_4}
\end{table}

\begin{table*} \centering
\centering \caption{Results on synthetic data (image $I_2$).}
\begin{tabular}{|c|c|c|c|c|c|c|c|c|}
  \cline{2-9}
\multicolumn{1}{c|}{}   & \multicolumn{4}{c|}{RMSE (classes) $(\times 10^{-2})$} & \multirow{2}{*}{RMSE}  & \multirow{2}{*}{RE} & \multirow{2}{*}{SAM} & \multirow{2}{*}{Time} \\
\cline{2-5} \multicolumn{1}{c|}{}   & $\mathcal{C}_1$  & $\mathcal{C}_2$ & $\mathcal{C}_3$ & $\mathcal{C}_4$  & \multirow{2}{*}{$(\times 10^{-2})$}   &  \multirow{2}{*}{$(\times 10^{-2})$} & \multirow{2}{*}{$(\times 10^{-2})$} & \multirow{2}{*}{(s)} \\
\multicolumn{1}{c|}{}   & LMM  & RCA-NL & GBM & PPNMM  &    &   &   & \\
\hline    FCLS     &  $10.24$        & $44.72$         & $15.48$        & $23.98$       &  $24.76 $        & $15.74 $       & $10.64 $       &  \blue{$1.7 $} \\
\hline SUNSAL-CLS  &  $3.84$         & $33.81$         & $5.68$         & $8.45$        &  $16.55 $        & $4.17 $        & $7.57 $        &  \red{$0.07 $} \\
\hline    SKhype   &  $1.67$         & $11.92$         & \blue{$2.21$}  & \blue{$2.81$} &  $5.87 $         & $- $           & $- $           & $547 $ \\
\hline    RCA-MCMC &  $5.87$         & \red{$6.29$}    & $5.44$         & $3.93$        &  \blue{$5.66 $}  & $- $           & $- $           & $9009$ \\
\hline    AEB      &  $53.33$        & $22.5$          & $49.2$         & $41.8$        &  $45.72$         & $3.05$         & $6.46$         & $1732$ \\
\hline    CDA-NL   &  \red{$1.62$}   & \blue{$7.27$}   & \red{$2.16$}   & $2.89$        &  \red{$3.86 $}   & \red{$2.86 $}  & \red{$6.16 $}  & $430$ \\
\hline    CDA-EV   &  $4.84$         & $33.27$         & $7.12$         & $9.01$        &  $16.59$         & $3.34 $        & $6.64 $        & $160 $ \\
\hline    CDA-ME   &  \blue{$1.63$}  & $13.55$         & $2.27$         & \red{$2.76$}  &  $6.61 $         & \blue{$2.89 $} & \blue{$6.17 $} & $37 $ \\
  \hline
\end{tabular}
\label{tab:Results_ImNL}
\end{table*}

\subsection{Robustness of the proposed algorithms} \label{subsec:Robustness_of_the_proposed_algorithm}
This section evaluates the robustness of the proposed CDA algorithms when varying the parameters $(R, L, N, \textrm{SNR})$. Synthetic images have been generated according to the proposed RCA models (NL, EV and ME)  with the following configurations: 
\begin{itemize}
\item The illumination coefficient increases
linearly from the left of the image to its right in the
interval [0.9; 1.15]. 
\item The abundances are uniformly distributed in the simplex.
\item The NL coefficient is fixed to $\epsilon^2 = 0.1$ in \eqref{eqt:priorGam}, the EV coefficient is fixed to  $\alpha^2 = 0.005, \forall r,i,j$ in \eqref{eqt:priorKcond}, and the ME coefficient is fixed to  $\epsilon^2 = 0.002$ in \eqref{eqt:priorD}. 
\end{itemize} 
Each CDA algorithm has been run on its corresponding image, i.e., CDA-NL for the NL image, CDA-EV for the EV image and CDA-ME for the ME image. Table \ref{tab:Results_Synth_L_N_SNR_R} reports the obtained RMSE results when varying $(R, L, N, \textrm{SNR})$. When not varying, the parameters are fixed to  $L=207$ bands, $N=10^4$ pixels, $R=3$ endmembers and SNR$=25$ dB. Table \ref{tab:Results_Synth_L_N_SNR_R} shows that the performance decreases when increasing the number of endmembers $R$. This result is similar to \cite{Halimi_TIP2015,Altmann2014b}, since increasing $R$ leads to an increase in the number of the estimated parameters which reduces the RMSE performance. In addition, Table \ref{tab:Results_Synth_L_N_SNR_R}  highlights   the decrease in performance  when reducing the observations either by reducing the number of bands $\#L$ or the number of pixels $\#N$. Note, however, that the results are still acceptable for $L\geq52$ bands and $N\geq 625$ pixels. As expected, the obtained results show that the higher the SNR, the better the RMSE performance is. These  results are in good agreement with the literature and confirm the good behavior of the CDA algorithms for different parameter configurations.  Note finally that more results are provided in the technical report \cite{HalimiTR2016_RCA} and are not shown here for brevity.
\begin{table}[h] \centering
\centering \caption{RMSE of the proposed CDA algorithms when varying $L$, $N$, $R$ and SNR. }
\begin{tabular}{|c|c|c|c|c|}
\cline{3-5} 
\cline{3-5} \multicolumn{2}{c|}{}    & CDA-NL         &  CDA-EV         & CDA-ME  \\
\hline
 \multirow{3}{*}{Varying $L$}        & $L=52$      &  $11.08$ & $6.34$ & $8.22$   \\
\cline{2-5}                          & $L=104$     &  $9.04$  & $4.61$ & $5.89$   \\
\cline{2-5}                          & $L=207$     &  $7.58$  & $3.67$ & $4.26$  \\ 
\hline
\hline 
 \multirow{4}{*}{Varying $N$}        & $N=169$     &  $21.00$ & $9.93$ & $9.52$   \\
\cline{2-5}                          & $N=625$     &  $7.71$  & $4.26$ & $4.32$  \\
\cline{2-5}                          & $N=2500$    &  $7.71$  & $3.74$ & $4.29$   \\
\cline{2-5}                          & $N=10000$   &  $7.58$  & $3.67$ & $4.26$  \\ 
\hline
\hline 
 \multirow{2}{*}{SNR=$15$}           & $R=3$     &  $14.52$ & $8.09$ & $9.05$  \\
\cline{2-5}                          & $R=6$     &  $19.21$ & $9.67$ & $11.11$    \\  
\hline
\hline 
 \multirow{2}{*}{SNR=$25$}           & $R=3$     &  $7.57$ & $3.62$ & $4.26$   \\
\cline{2-5}                          & $R=6$     &  $13.94$ & $5.95$ & $7.52$  \\   
\hline 
\end{tabular}
\label{tab:Results_Synth_L_N_SNR_R}
\end{table}

\section{Simulation results on real data} \label{sec:Simulation_results_on_real_data}
 
\begin{figure}[h!]
\centering \subfigure[Moffett image.]{\includegraphics[width=0.4\figwidth,height=3.6cm]{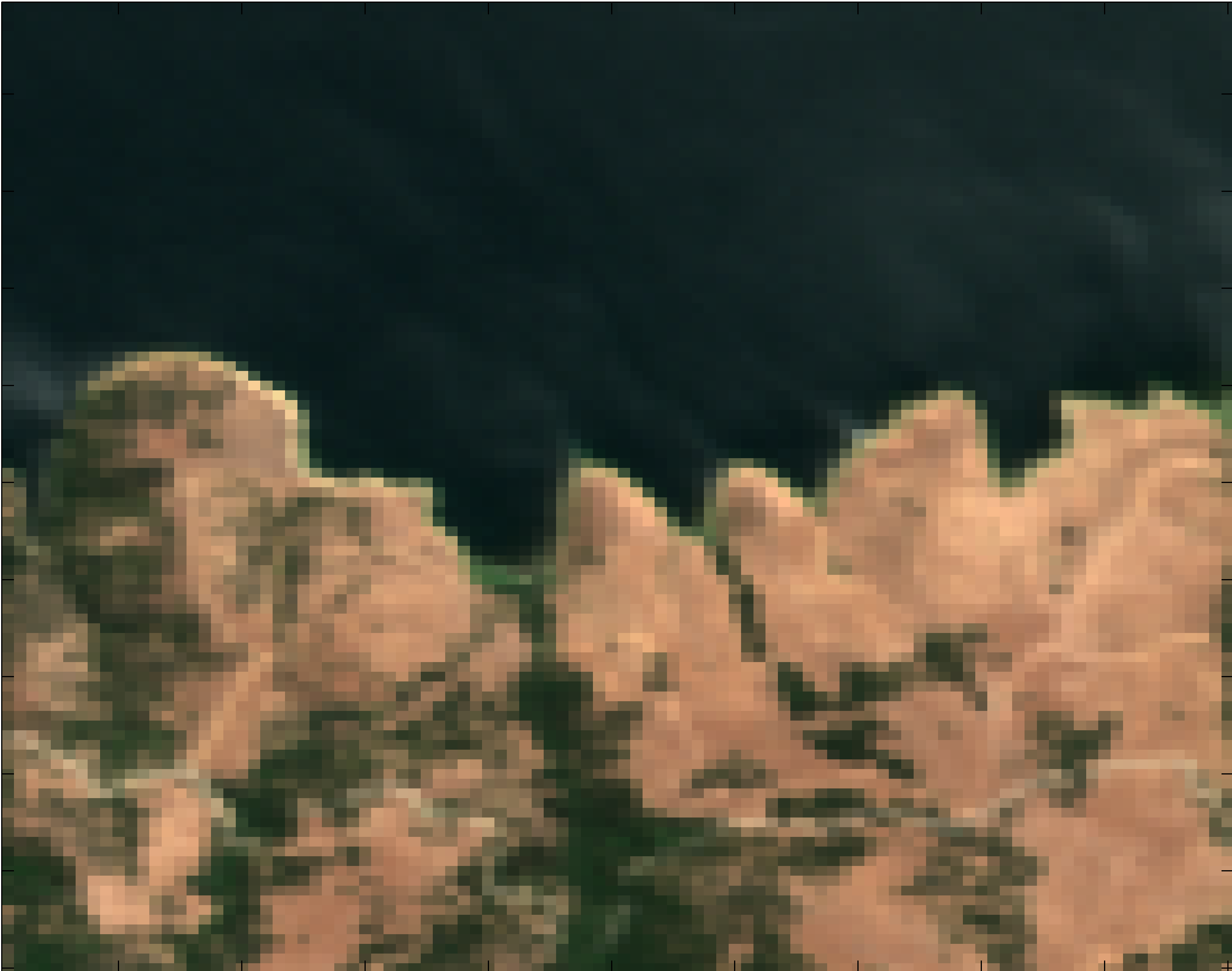}}\hspace{0.25cm}
\subfigure[Madonna image.]{\includegraphics[width=0.4\figwidth,height=3.6cm]{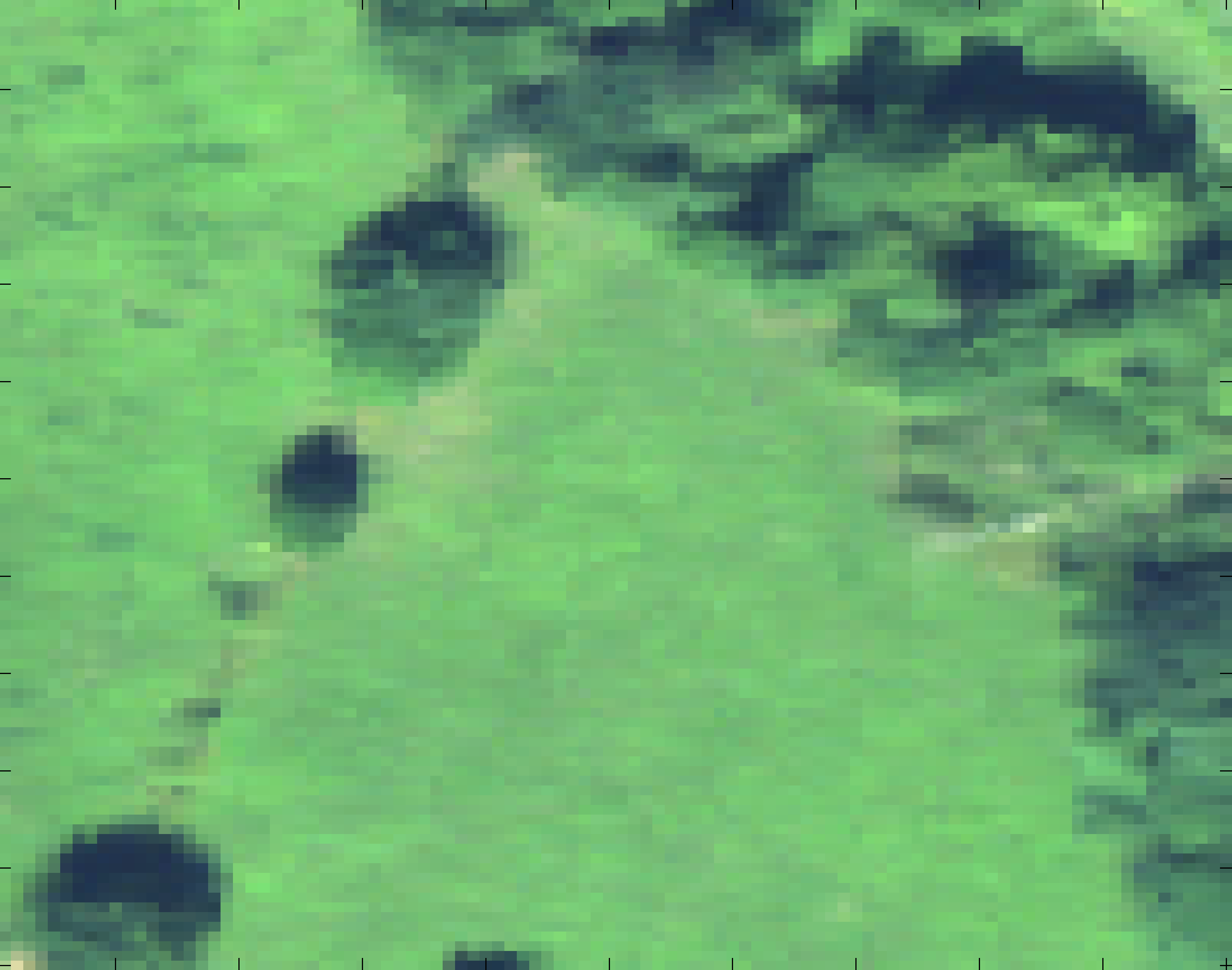}}
\caption{Real hyperspectral images. (Left) Moffett image, (right) Madonna image.} \label{fig:TIP_Real_images}
\end{figure}

This section illustrates the performance of the proposed algorithms when applied to two real hyperspectral images. The first hyperspectral image has received much attention
in the remote sensing community \cite{Halimi2011TGRS,Dobigeon2008}. This image was acquired over Moffett Field, CA, in 1997 by the AVIRIS. The considered dataset contains $100 \times 100$ pixels, $L=152$ spectral bands (after removing water absorption bands) acquired in the interval $0.4-2.5 \mu$m, has a spatial resolution of $100$m and is mainly composed of three components: water, soil, and vegetation (see Fig. \ref{fig:TIP_Real_images} (left)).  For this image,  three endmembers were extracted using VCA \cite{Nascimento2005}.
The second  image, denoted as Madonna, was acquired in $2010$ by the Hyspex hyperspectral scanner over Villelongue, France (00 03'W and 4257'N). The dataset contains $L=160$ spectral bands recorded from the visible to near infrared ($400-1000$nm) with a spatial resolution of $0.5$m  \cite{Sheeren2011}. It has already been studied in \cite{Halimi_TIP2015,Altmann2014b} and is mainly composed of forested areas (see Fig. \ref{fig:TIP_Real_images} (right)).
This image contains $100 \times 100$ pixels and is composed of $R=4$ components: tree, grass, soil and shadow (see Fig. \ref{fig:TIP_Real_images} (right)). The Bayesian UsLMM algorithm \cite{Dobigeon2009}  was used to estimate $R= 4$ endmembers.

\subsection{Unmixing performance} \label{subsec:Unmixing_performance}
\begin{figure}[h!]
\centering
\includegraphics[width=0.95\figwidth,height=12cm]{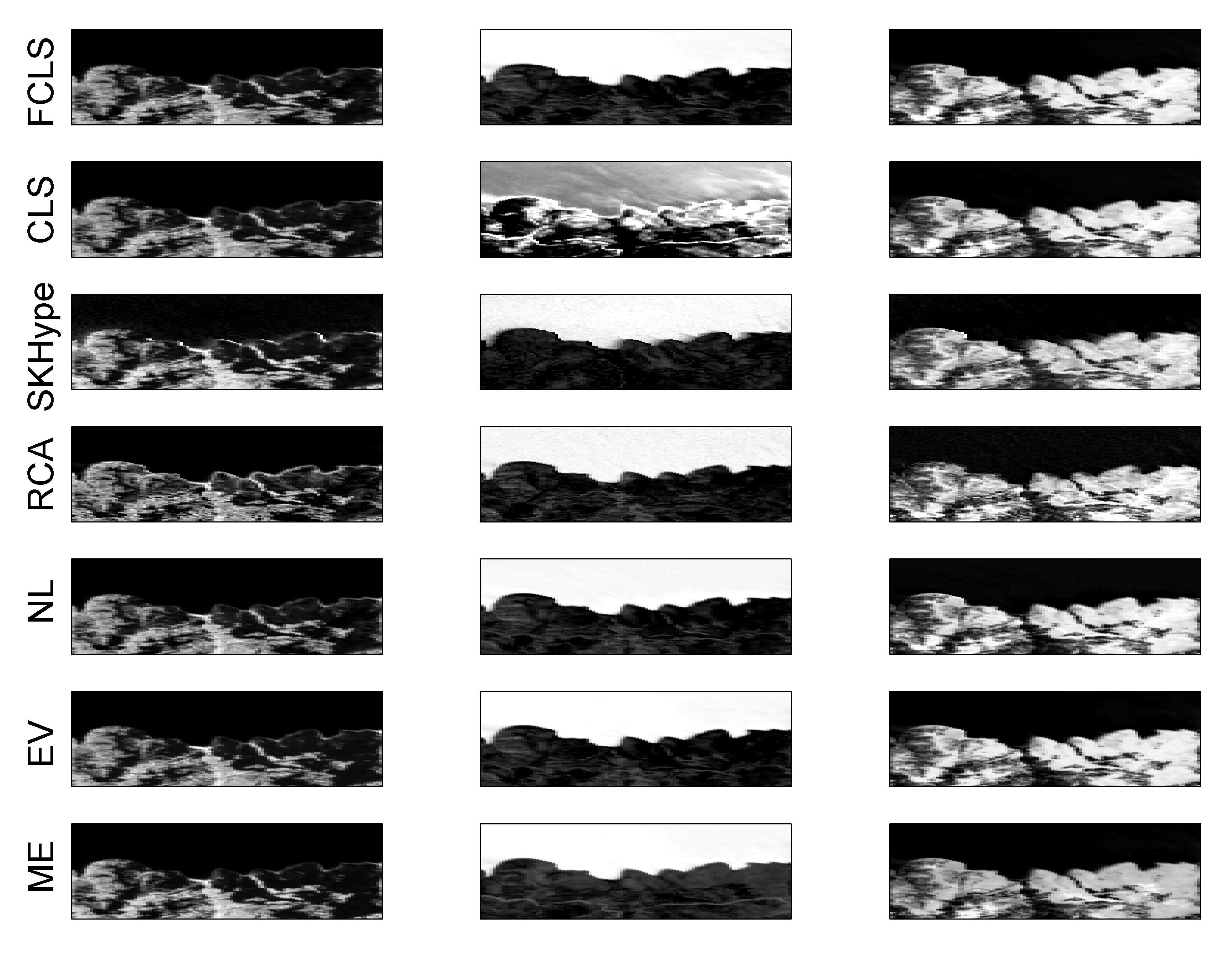}
\caption{Estimated abundance maps with different algorithms for the Moffett image.  (Left) vegetation, (middle) water, (right) soil.} \label{fig:Abundances_im5_Moffett_Large_HCT_CF}
\end{figure}

The abundances of each image have been estimated by the considered unmixing algorithms. Fig. \ref{fig:Abundances_im5_Moffett_Large_HCT_CF} shows the obtained results for the Moffett image. Note that a white (black) pixel indicates a large (small) proportion of the corresponding materials. Except SUNSAL-CLS, the considered algorithms show similar abundance maps. The behavior of SUNSAL-CLS is due to the presence of a high illumination variation for this image. In addition, some nonlinearity can be interpreted as an illumination variability (as already shown when analysing synthetic data) leading to high values for the parameter $\bsc$. This behavior will be further analysed in the next sections. Note that the algorithms estimated similar abundance maps for the Madonna image and for conciseness, the abundances maps are not displayed. The unmixing performance can also be compared by considering the reconstruction errors as shown in  Table \ref{tab:Results_ImReal}. Considering the Moffett image, the best RE and SAM results  are obtained by CDA-ME followed by CDA-EV and CDA-NL. These results suggest the presence of both EV and NL in the Moffett image which requires the use of a robust unmixing algorithm such as CDA-ME (a detailed study of NL and EV will be provided in the next sections). The best performance on the Madonna image are obtained by the CDA-EV algorithm, while we obtain good results when considering the algorithms including EV such as AEB, CDA-ME and SUNSAL-CLS. These results suggest the presence of a higher variability effect than nonlinearity in the Madonna image. Note finally that the proposed CDA algorithms show a reduced computational cost while providing many details regarding the physical effects corrupting the images as shown in the next sections.


\begin{table*} \centering
\centering \caption{Results on real images.}
\begin{tabular}{|c|c|c|c||c|c|c|}
  \cline{2-7}
\multicolumn{1}{c|}{}   & \multicolumn{3}{c|}{Moffett image} & \multicolumn{3}{|c|}{Madonna image}  \\
\cline{2-7}
\multicolumn{1}{c|}{} & RE & SAM & Time  & RE & SAM & Time \\
\cline{2-7}
\multicolumn{1}{c|}{} & $(\times 10^{-3})$ & $(\times 10^{-2})$ & (s)  & $(\times 10^{-3})$ & $(\times 10^{-2})$ & (s) \\
\hline    FCLS     &  $13.6$         & $12.7$          & \blue{$2$}     & $6.3$          &  $2.7 $        & \blue{$1$} \\
\hline SUNSAL-CLS  &  $6.4$          & $9.1$           & \red{$0.1$}    & \blue{$6.2$}   &  \blue{$2.6 $} & \red{$0.1 $} \\
\hline    SKhype   &  $-$            & $-$             & $241$          & $-$            &  $-  $         & $131$           \\
\hline    RCA-MCMC &  $-$            & $-$             & $11160$        & $-$            &  $-  $         & $13206$           \\
\hline    AEB      &  $9.7$          & $9.7$           & $324$         & \blue{$6.2$}   &  $2.6$         & $1841$    \\
\hline    CDA-NL   &  $5.7$          & $9.0$           & $364$         & $6.4$          &  $2.7$         & $1588$ \\
\hline    CDA-EV   &  \blue{$3.8$}   & \blue{$5.57$}   & $519$         & \red{$6.0$}    &  \red{$2.5$}   & $1277$        \\
\hline    CDA-ME   &  \red{$2.9$}    & \red{$3.25$}    & $259$          & \blue{$6.2$}   &  \blue{$2.6$}  & $91$  \\
  \hline
\end{tabular}
\label{tab:Results_ImReal}
\end{table*}


\subsection{Residual components} \label{subsec:Residual_components}
\begin{figure}[h!]
\centering
\includegraphics[width=0.95\figwidth,height=5.5cm]{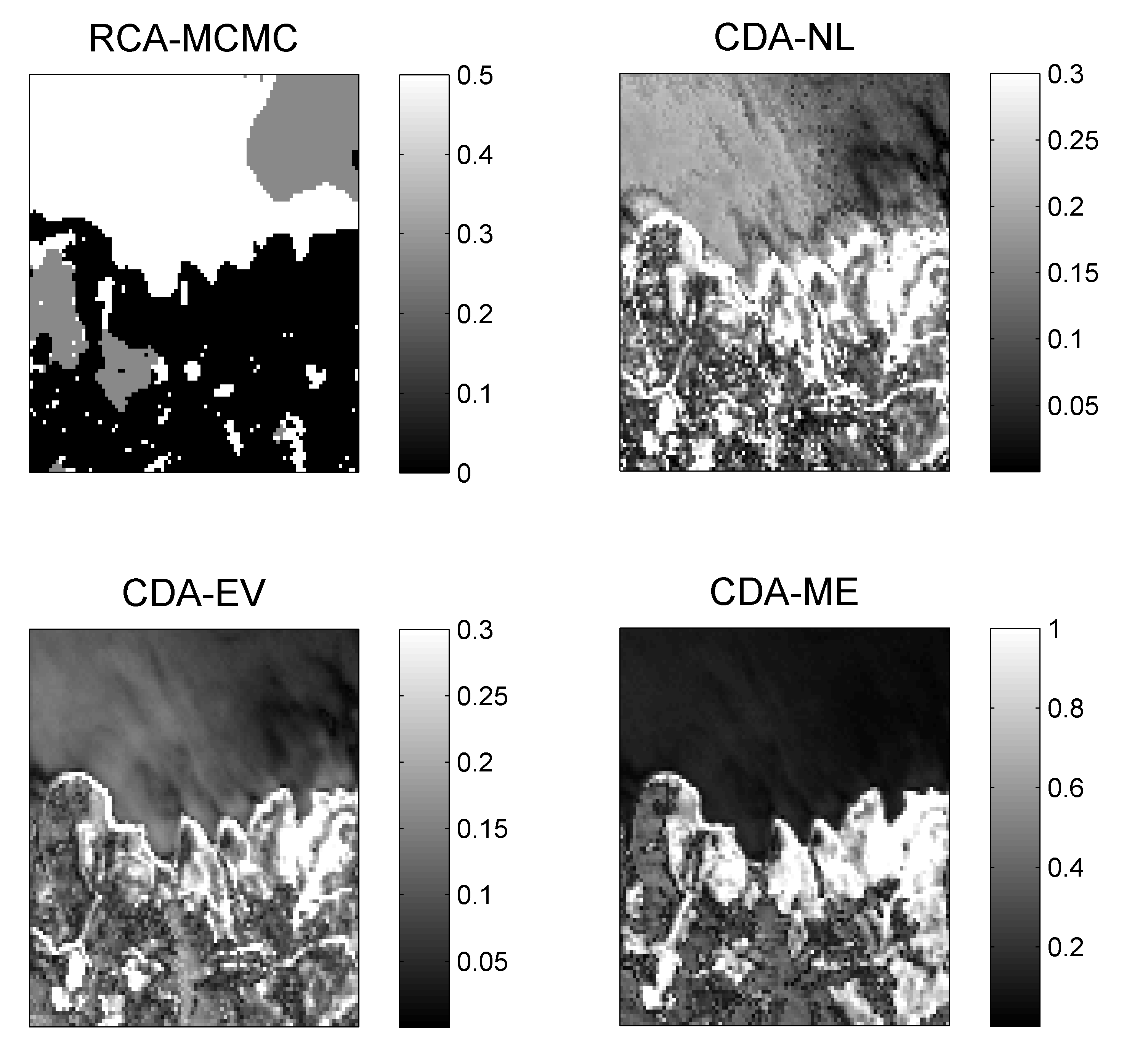}
\caption{Square root of the energies of the difference between the reconstructed signal and the linear model obtained with $||\hat{\bsy}_{i,j} - \bsM \hat{\bsa}_{i,j}||$ (for RCA-MCMC we show the estimated nonlinearity coefficients) for the Moffett image.} \label{fig:DiffLMM_im5_Moffett_Large_HCT_CF}
\end{figure}

This section analyses the obtained residual energies for the considered real images. Fig. \ref{fig:DiffLMM_im5_Moffett_Large_HCT_CF} shows the energies of the difference between the reconstructed signal and the linear model (i.e.,  $||\hat{\bsy}_{i,j} - \bsM \hat{\bsa}_{i,j}||$) for the Moffett image.  For RCA-MCMC, the reconstructed spectrum is not available and we only show the estimated nonlinearity levels. Fig. \ref{fig:DiffLMM_im5_Moffett_Large_HCT_CF} shows a good agreement between the results of the proposed CDA algorithms where mismodelling effects are mainly located in the coastal zone, and in the right area (composed of $\approx 90\%$ soil and $\approx 10\%$ vegetation). Similarly to RCA-MCMC, the three CDA algorithms show some mismodelling effects in the left zone of the water area. However, the energy location shows some differences between RCA-MCMC and CDA-algorithms mainly because of the presence of both NL and EV in the Moffett scene as shown in the previous section (see Table \ref{tab:Results_ImReal}). Indeed, all CDA algorithms account for EV illumination effect while RCA-MCMC only account for NL effect.

\begin{figure}[h!]
\centering
\includegraphics[width=0.95\figwidth,height=5.5cm]{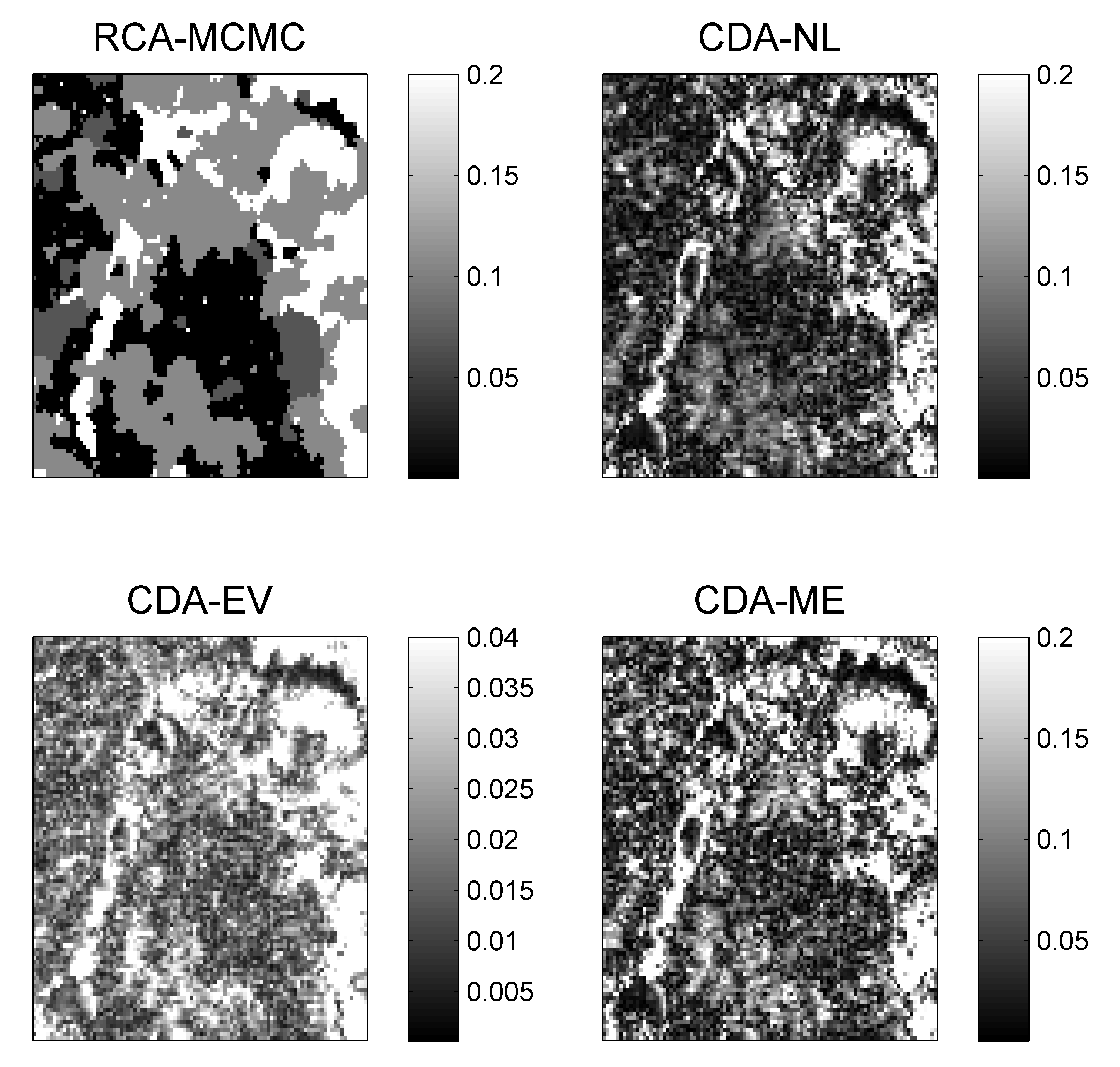}
\caption{Square root of the energies of the difference between the reconstructed signal and the linear model  obtained with $||\hat{\bsy}_{i,j} - \bsM \hat{\bsa}_{i,j}||$ (for RCA-MCMC we show the estimated nonlinearity coefficients) for the Madonna image.} \label{fig:DiffLMM_im2_MadonaNL_Large_HCT_CF}
\end{figure}

The same energies were displayed when considering the Madonna image in Fig. \ref{fig:DiffLMM_im2_MadonaNL_Large_HCT_CF}. This figure shows a good agreement between the results of RCA-MCMC and CDA algorithms where the high deviation from the linear model is mainly located in the tree and shadow areas (as in \cite{Altmann2014b}).
This makes sense since multiscattering effects (i.e., nonlinear interactions) are generally located in these areas.
 Note finally that for both CDA-NL and CDA-ME, the displayed maps in Figs. \ref{fig:DiffLMM_im5_Moffett_Large_HCT_CF} and  \ref{fig:DiffLMM_im2_MadonaNL_Large_HCT_CF} include the effect of the illumination variation (introduced by $c_{i,j}$) and that of the residual term $\bphi$. These two terms are studied separately in the next section.
   
\subsection{Nonlinearity and illumination} \label{subsec:Nonlinearity}
\begin{figure}[h!]
\centering
\includegraphics[width=0.95\figwidth,height=5.5cm]{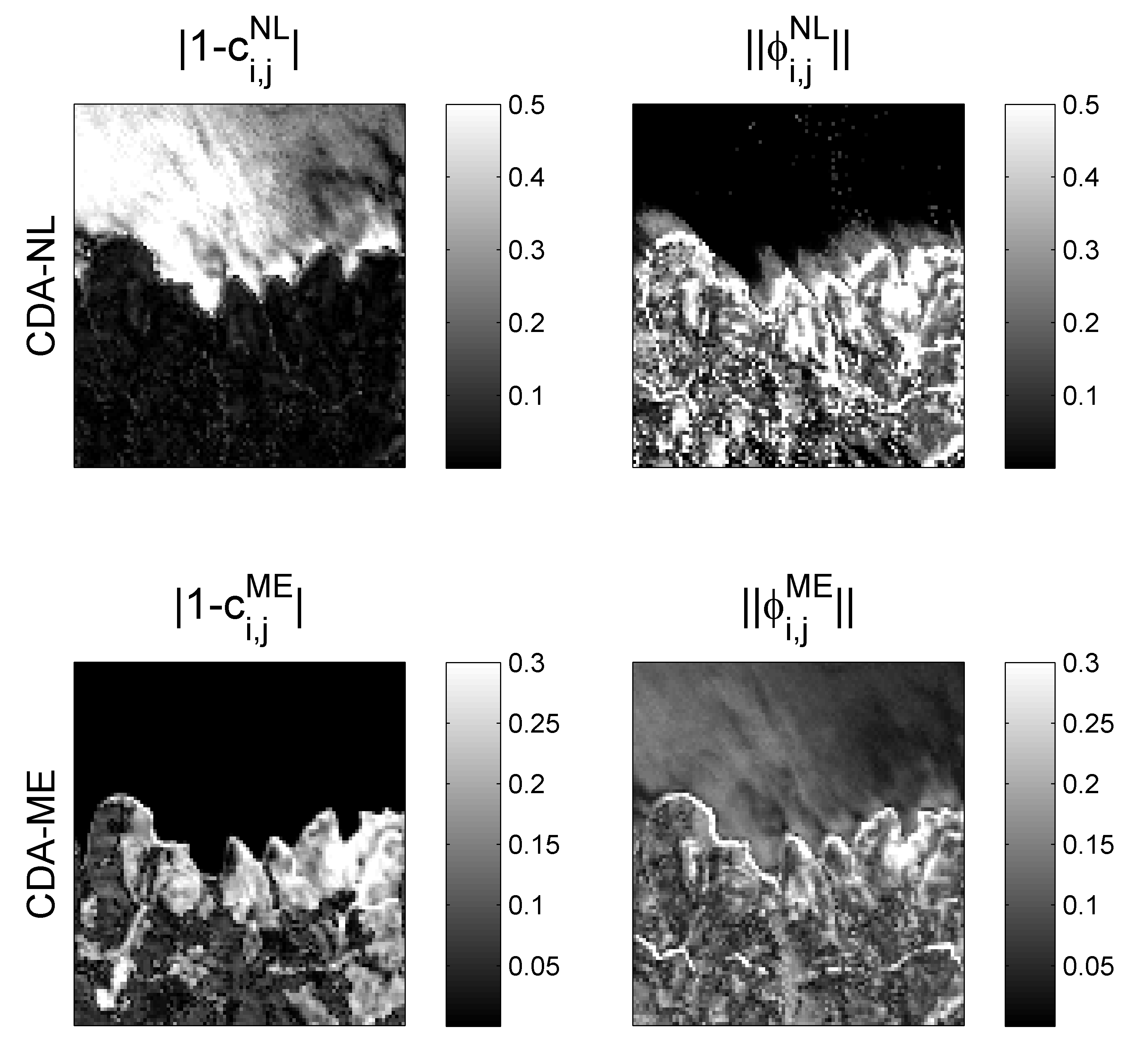}
\caption{Residual maps obtained with CDA-NL and CDA-ME algorithms for the Moffett image. (Left) estimated illumination variation $|1-c_{i,j}|$, (right) square root of the energies of the residual terms $\left\|\phi_{i,j}\right\|$.} \label{fig:C_Phi_im5_Moffett_Large_HCT_CF}
\end{figure}

As previously noticed, the Moffett image includes both NL and EV effects. Fig. \ref{fig:C_Phi_im5_Moffett_Large_HCT_CF} separates the residual energies into two components: the first one is related to illumination parameter $c$ and the second one to the term $\bphi$. Considering CDA-NL, it can be seen that the nonlinearity captured by $\bphi^{NL}$ is mainly located in the coastal region and areas comprising multiple physical components (as in \cite{Halimi2011TGRS}). This algorithm also captures a high variation in illumination in the left of the water area which has also been captured by RCA-MCMC (see Fig. \ref{fig:DiffLMM_im5_Moffett_Large_HCT_CF}), however, it has been interpreted as a nonlinear effect. These results show that there is a close relation between the effects of illumination variation and nonlinearity. This remark can be further justified by considering the results of CDA-ME which captures both EV and NL effects. Fig. \ref{fig:C_Phi_im5_Moffett_Large_HCT_CF} (left-bottom) shows that most of the NL effect can be interpreted as a variation of illumination while the remaining effects can be approximated by a residual smooth term $\bphi^{ME}$. Therefore, in presence of nonlinearity, a great unmixing improvement can be obtained by allowing an illumination variation, which also justifies the great improvement of SUNSAL-CLS with respect to FCLS for the Moffett image (see Table \ref{tab:Results_ImReal}).

\begin{figure}[h!]
\centering
\includegraphics[width=0.95\figwidth,height=5.5cm]{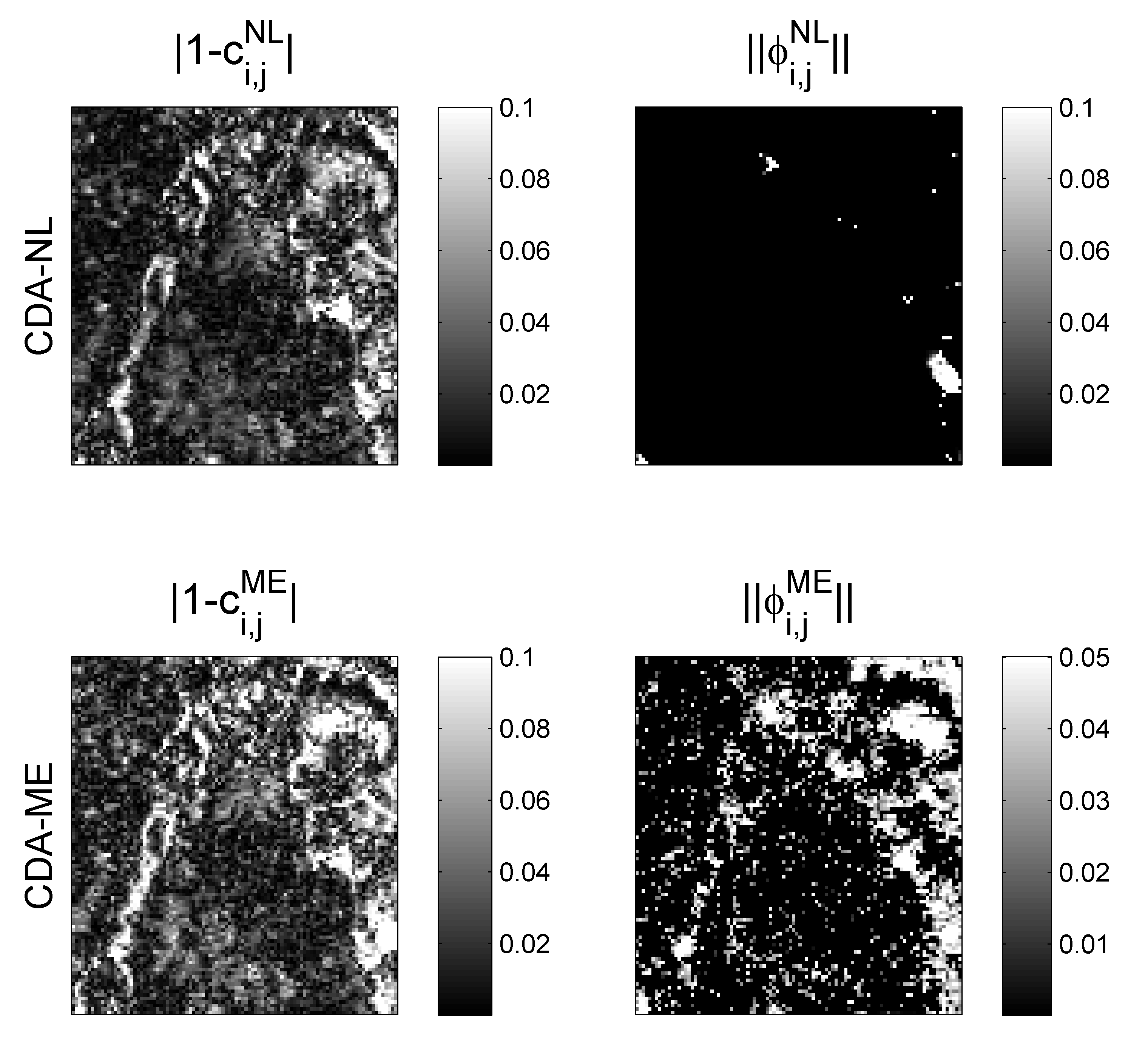}
\caption{Residual maps obtained with CDA-NL and CDA-ME algorithms for the Madonna image.  (Left) estimated illumination variation $|1-c_{i,j}|$, (right) square root of the energies of the residual terms $\left\|\phi_{i,j}\right\|$.} \label{fig:C_Phi_im2_MadonaNL_Large_HCT_CF}
\end{figure}

Considering Madonna image, Fig. \ref{fig:C_Phi_im2_MadonaNL_Large_HCT_CF} (top) shows a reduced NL effect and most of the residuals can be interpreted as a variation in illumination. Similarly to CDA-NL, CDA-ME captures an illumination variation that is mainly located in the region of trees. In addition, Fig. \ref{fig:C_Phi_im2_MadonaNL_Large_HCT_CF} (right-bottom) shows that CDA-ME also captures the shape EV as for CDA-EV. Indeed, CDA-ME is an intermediate algorithm that can deal with illumination variation, NL and EV effects.
To conclude, this section shows that most of the NL effect can be captured by varying the illumination coefficient. Moreover, CDA algorithms capture similar spatial residual effects but give a different interpretation depending on the considered mixture model. Most of these residuals appear in region of intersection between the physical elements and in presence of vegetation such as trees.  Finally, CDA-ME is very flexible and can capture different kind of mismodelling effects.
 
\subsection{Endmember variability} \label{subsec:Endmember_variability}
\begin{figure}[h!]
\centering
\includegraphics[width=0.95\figwidth,height=3cm]{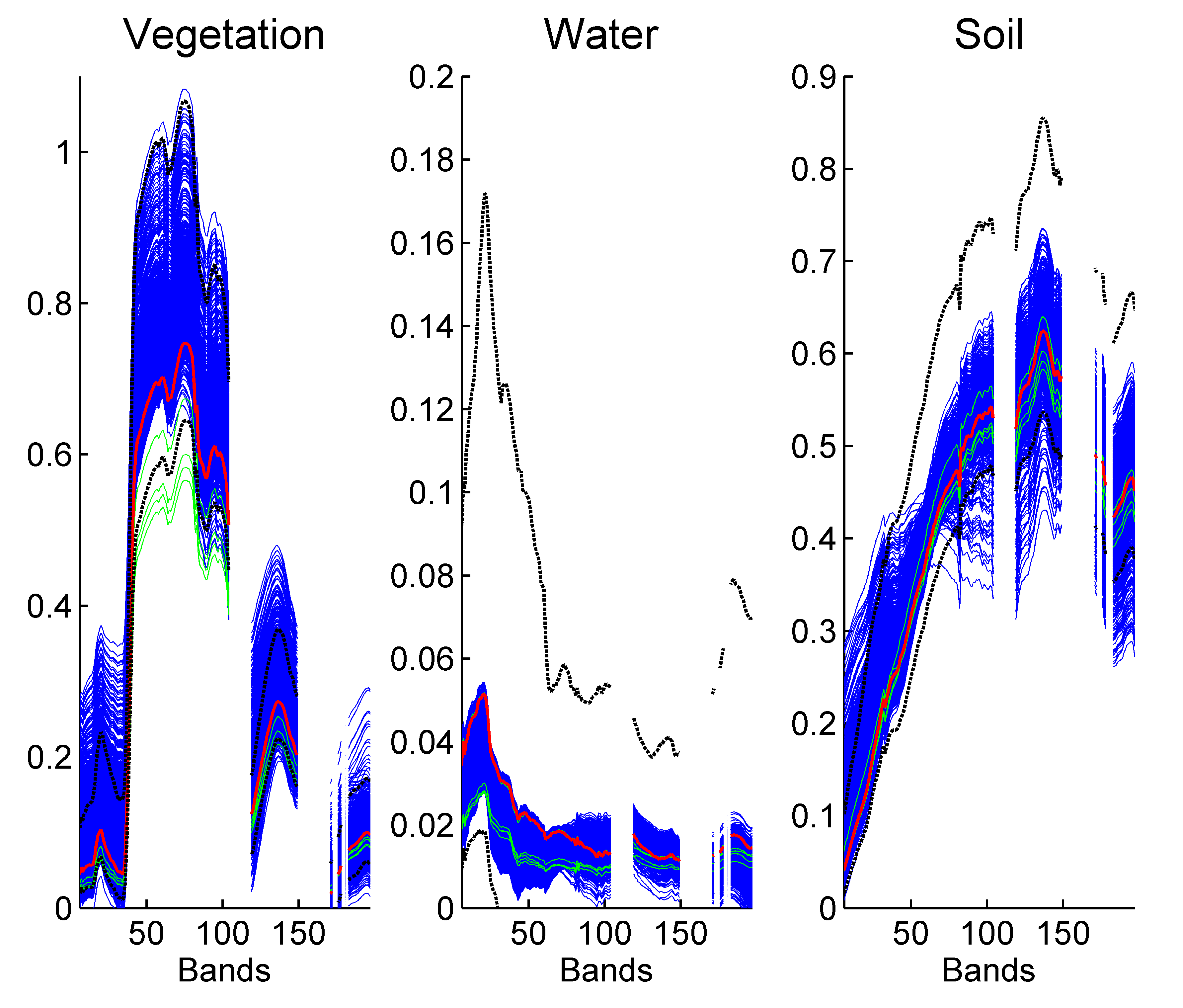}
\caption{The estimated $R = 3$ endmembers of the Moffett image with VCA (continuous red lines), AEB (continuous green lines), CDA-EV (continuous blue lines) and the  interval of spectra with CDA-ME (dashed black lines). } \label{fig:Endmembers_im5_Moffett_Large_HCT_CF}
\end{figure}

\begin{figure}[h!]
\centering
\includegraphics[width=0.95\figwidth,height=3cm]{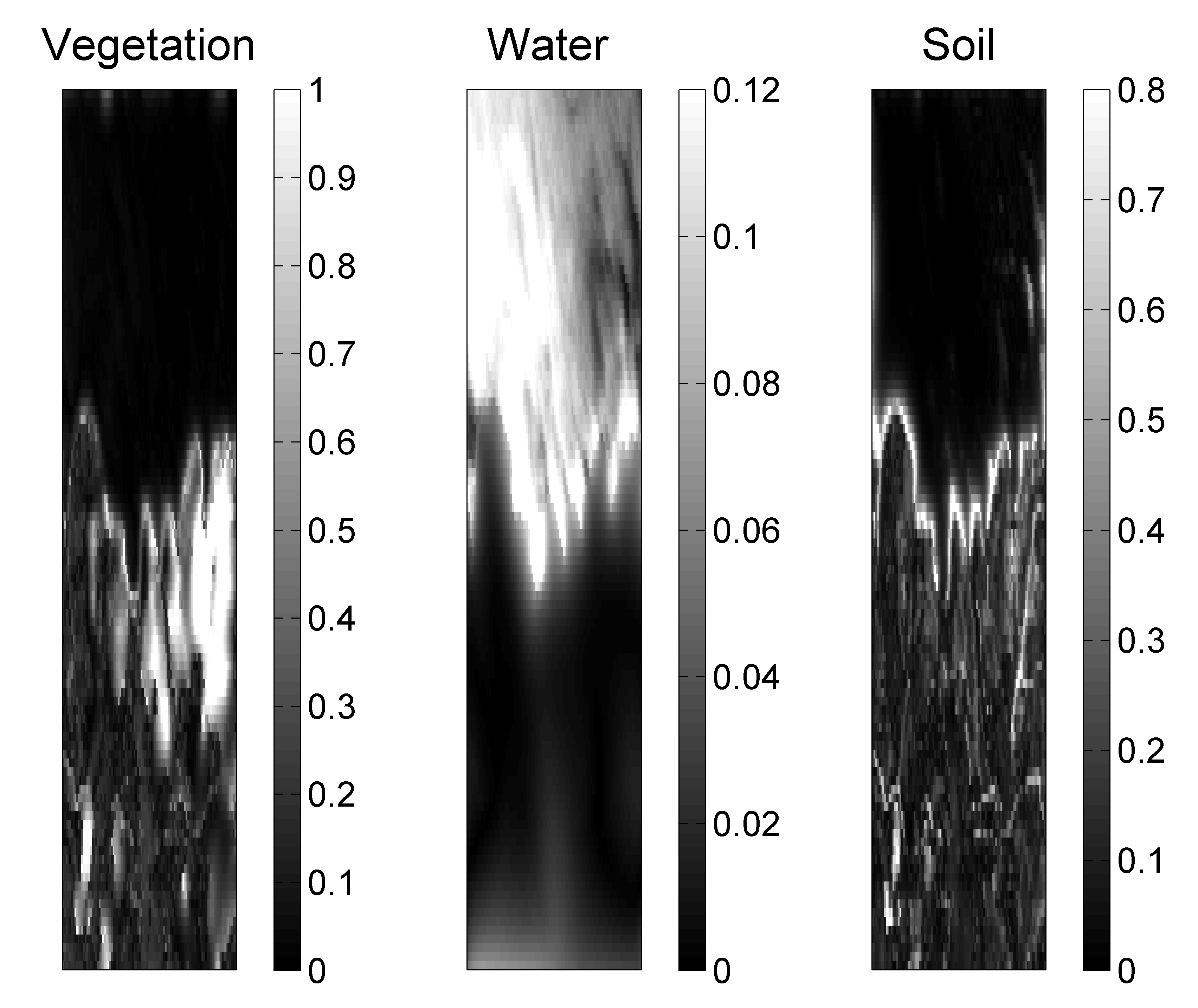}
\caption{Estimated spatial variability maps with CDA-EV for the Moffett image. The $\#r$th map is obtained by computing $\left\|\bsk_{i,j,r}\right\|$ for each pixel.} \label{fig:Maps_Variability_im5_Moffett_Large_HCT_CF}
\end{figure}

\begin{figure}[h!]
\centering
\includegraphics[width=0.95\figwidth,height=5.5cm]{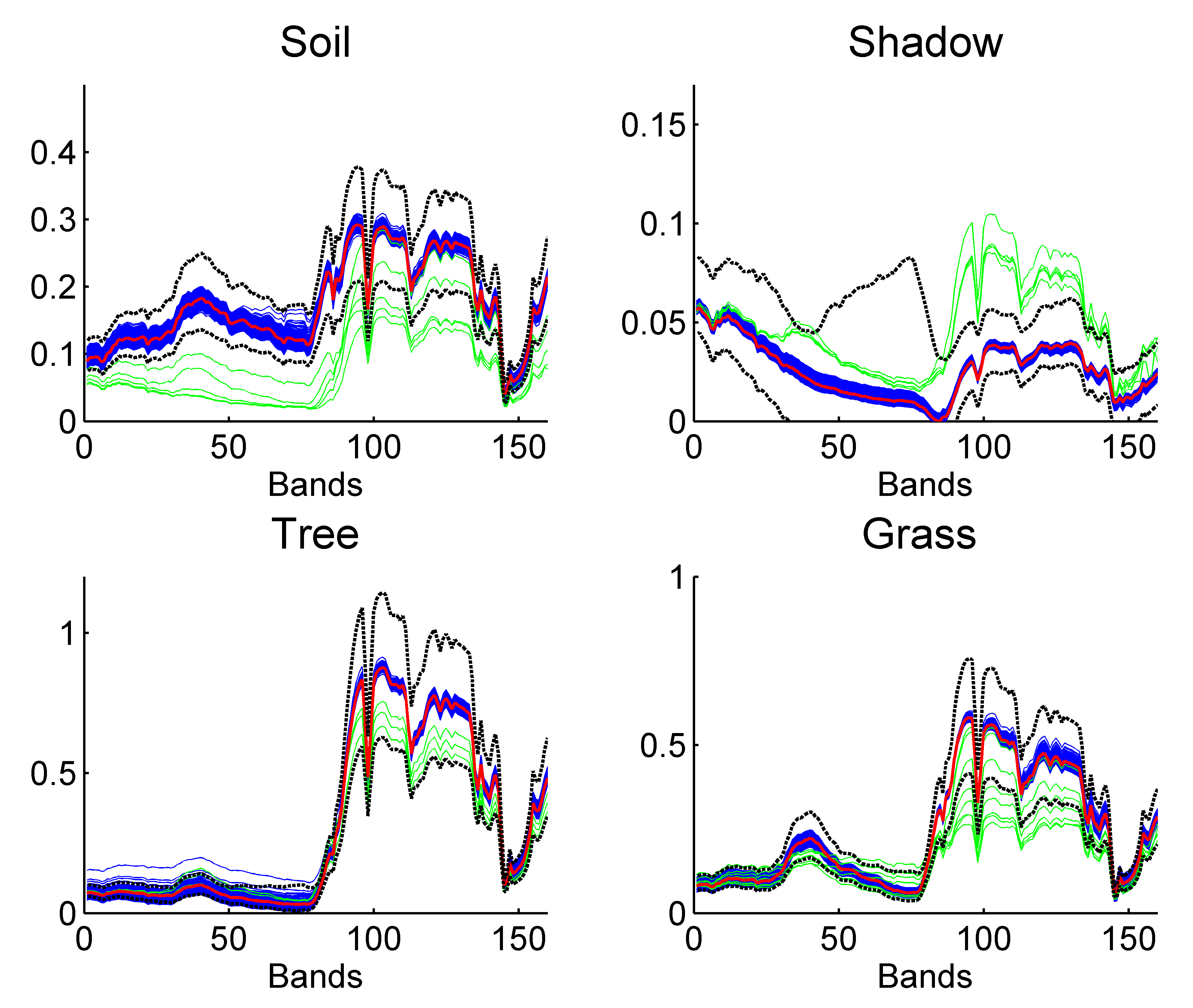}
\caption{The estimated $R = 4$ endmembers of the Madonna image with UsLMM (continuous red lines), AEB (continuous green lines), CDA-EV (continuous blue lines) and the  interval of spectra with CDA-ME (dashed black lines). } \label{fig:Endmembers_im1_MadonaEV_Large_HCT}
\end{figure}

\begin{figure}[h!]
\centering
\includegraphics[width=0.95\figwidth,height=5.5cm]{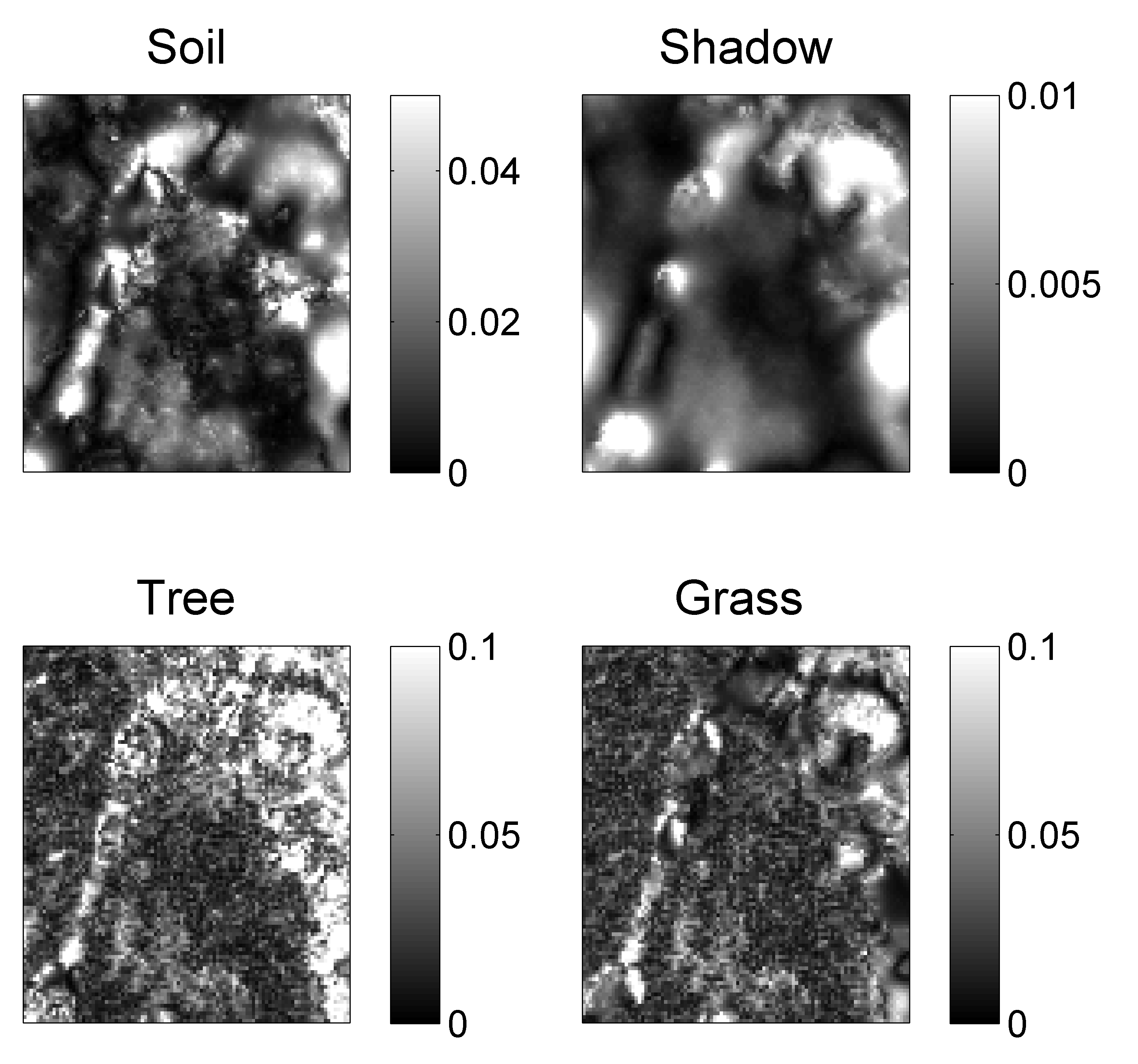}
\caption{Estimated spatial variability maps with CDA-EV for the Madonna image. The $\#r$th map is obtained by computing $\left\|\bsk_{i,j,r}\right\|$ for each pixel.} \label{fig:Maps_Variability_im1_MadonaEV_Large_HCT_CF}
\end{figure}

CDA-EV estimates a set of endmember spectra for each  pixel to account for EV. Figs. \ref{fig:Endmembers_im5_Moffett_Large_HCT_CF} and \ref{fig:Endmembers_im1_MadonaEV_Large_HCT}  compare the estimated spectra with those extracted with AEB, VCA or UsLMM. These figures also show the interval of endmembers obtained with CDA-ME when interpreting the mismodelling term $\bsd$ as to be due to EV. Overall, the estimated spectra are in good agreement with the state-of-the-art algorithms especially for the Moffett image. These figures  show that CDA-ME captures effects that are not only due to EV since negative spectra may appear. Figs. \ref{fig:Maps_Variability_im5_Moffett_Large_HCT_CF} and \ref{fig:Maps_Variability_im1_MadonaEV_Large_HCT_CF} show the spatial repartition of the captured EV by CDA-EV. For both images, the highest EV is obtained for vegetation (tree and grass) and for regions with multiple components.
For the Moffett image, and as an example, the water variability is located in the left area of the water zone while the soil variability is concentrated in the coastal area. For the Madonna image, the tree variability appears in the forest region while the shadow variability is located in the shadowed regions. These results show the ability of CDA-EV to provide both the spectral behavior and the spatial location of EV.

We conclude this section by providing some comments regarding the proposed algorithms and their use in practice. It has been shown in this section that CDA-ME can capture both NL and EV in presence of vegetation, terrain relief or multiple physical components. In addition to this flexibility, CDA-ME is generally fast, which encourages its use as a first step to analyze a scene. The obtained results can then be analyzed to highlight the presence of EV or NL. As a second step,  CDA-NL or  CDA-EV can be used to focus on the phenomenon of interest. CDA-EV is designed to highlight the sensitive spectral bands as well as the shape of the spectral variation  and the  regions of spatial variation of each materiel.  CDA-NL is designed to highlight the region of interaction of the physical components as well as the active bilinear components. Finally, we provide Table \ref{tab:Models} that summarizes some characteristics and differences between the proposed models.
\renewcommand{\arraystretch}{1.2}
\begin{table}[h] \centering
\centering \caption{Characteristics of the proposed models. ``Pos.'' stands for positivity, ``Illumin.'' for Illumination, ``correl.'' for correlation, (++) best results, and (+) good results.  }
\begin{tabular}{|c|c|c|c|c|c|}
\cline{2-6} \multicolumn{1}{c|}{}      & \multicolumn{3}{c|}{Residuals}  &  Illumin. &  \multirow{2}{*}{Time} \\ 
\cline{2-4} \multicolumn{1}{c|}{}      & Pos. & Spatial    & Spectral    &  coeff. c & \\ 
\hline  \multirow{2}{*}{NL}    & \multirow{2}{*}{$\checkmark$}    &  Correl.    & \multirow{2}{*}{\text{\sffamily X}}  & \multirow{2}{*}{$\checkmark$} &  + \\ 
                               &    & energies   &  &   & \\  
\hline  \multirow{2}{*}{EV}    & \multirow{2}{*}{\text{\sffamily X}}    &  Correl.    & Correl.  & \multirow{2}{*}{\text{\sffamily X}} &  + \\ 
                               &    & values   &  values &   & \\		 
\hline  \multirow{2}{*}{ME}    & \multirow{2}{*}{\text{\sffamily X}}    &  Correl.    & Correl.  & \multirow{2}{*}{$\checkmark$} &  ++ \\ 
                               &    & energies   &  values &   & \\		 
\hline
\end{tabular}
\label{tab:Models}
\end{table}

\section{Conclusions} \label{sec:Conclusions} 
This paper introduced three hyperspectral mixture models and their associated Bayesian algorithms
for supervised hyperspectral unmixing. The three models were introduced under a general formulation that can be adapted to account for nonlinearity, endmember variability or mismodelling effects. A hierarchical Bayesian model was proposed for each model to introduce the known constraints on their parameters. Those parameters were estimated using a coordinate descent algorithm that showed a reduced computational cost when compared to state-of-the-art algorithms. The proposed algorithms showed good performance when processing synthetic data generated with the linear model or other more sophisticated models. Results on real data confirmed the good performance of the proposed algorithms and showed their ability to extract different features in the observed scenes. These results showed that most of the nonlinearity can be interpreted by a variation in illumination, that endmember variability is mainly located in vegetation areas, and that residual effects (such as nonlinearity or endmember variability) appears in presence of multiple physical components.
Future work includes the estimation of the hyperparameters associated with the proposed algorithms, since they control the degree of smoothness of the obtained results (see for instance Figs. \ref{fig:Maps_Variability_im5_Moffett_Large_HCT_CF} and \ref{fig:Maps_Variability_im1_MadonaEV_Large_HCT_CF}). It also includes the estimation of the endmember  number using a model selection criteria such as the Akaike information criterion (AIC) \cite{Akaike1974}, the minimum description length (MDL) \cite{Rissanen1978}, or other methods such as \cite{Bioucas2008,HalimiTGRS2016}. 
Adapting the proposed algorithms to include a  dimension reduction step is an interesting issue which should further reduce their computational cost. Finally, detecting the presence of endmember variability and nonlinearity using hypothesis tests is also an interesting issue which would deserve to be investigated. 

\vspace{-0.2cm}
\appendices
\appendix[Conditional distributions] \label{app:Conditional_distributions}
This appendix provides the conditional distributions of the estimated parameters where $\bsS_{i,j}$ and $\bphi_{i,j}$ depend on the considered observation model (NL, EV or ME). The conditional distributions are given by
\begin{equation} \left\lbrace
\begin{tabular}{l}
$\bsa_{i,j} | \bThe_{\backslash \bsa_{i,j}}   \sim \calN_{\calS}\left(\bmu_{i,j}^{a}, \bLam_{i,j}^{a} \right)$\\
$\bgam_{i,j} | \bThe_{\backslash \bgam_{i,j}}   \sim \calN_{{(\dsR+)}^D} \left(\bmu_{i,j}^{\gamma}, \bLam_{i,j}^{\gamma} \right)$\\
$\bsk_{r,i,j} | \bThe_{\backslash \bsk_{r,i,j}}  \sim  \calN \left( \bmu_{r,i,j}^{k}, \bLam_{r,i,j}^{k}\right)$\\
$\bsd_{i,j} | \bThe_{\backslash \bgam_{i,j}}   \sim \calN  \left(\bmu_{i,j}^{d}, \bLam_{i,j}^{d} \right)$\\
$c_{i,j}^{ME} | \bThe_{\backslash c_{i,j}}   \sim \calN \left(\bmu_{i,j}^{c}, \bLam_{i,j}^{c} \right)$  \\
$f\left(c_{i,j}^{NL} | \bThe_{\backslash c_{i,j}} \right)\propto
\exp{\left(\sum_{m=0}^{4}{x_m c_{i,j}^{m}} \right)}$\\ 
$\sigma^2_{\ell} | \bThe_{\backslash \sigma^2_{\ell} }   \sim \calI \calG \left(\varphi_{\ell}+\frac{N}{2},  \psi_{\ell} +  \sum_{i,j}{\frac{\left[\bsy_{i,j}(\ell)- \bsS_{i,j}(\ell) \bsa_{i,j} - \bphi_{i,j}(\ell)\right]^2}{2}} \right)$\\
NL:  $\epsilon^2_{i,j} | \bThe_{\backslash \epsilon^2_{i,j} }   \sim \calI \calG \left(4 \zeta + \frac{D}{2},  \frac{\bgam_{i,j}^{\top}\bgam_{i,j}}{2} +  4 \zeta \rho_{1,i,j}(\bsw)  \right)$ \\
ME: $\epsilon^2_{i,j} | \bThe_{\backslash \epsilon^2_{i,j} }   \sim \calI \calG \left(4 \zeta + \frac{L}{2},  \frac{\bsd_{i,j}^{\top} \bsH^{-1} \bsd_{i,j}}{2} +  4 \zeta \rho_{1,i,j}(\bsw)  \right)$
\\
\end{tabular}\right.
\end{equation}
where 
\begin{equation} \left\lbrace
\begin{tabular}{l}
$\bLam_{i,j}^{a} = \left( \bsS_{i,j}^{\top}  \bSig^{-1} \bsS_{i,j}  \right)^{-1}$ \\
$\bmu_{i,j}^{a} = \bLam_{i,j}^{a} \bsS_{i,j}^{\top} \bSig^{-1}  \left(\bsy_{i,j}-\bphi_{i,j}\right)$\\
$ \bLam^{\gamma}_{i,j} =  \left(c_{i,j}^{4} \bsQ^{\top}  \bSig^{-1} \bsQ + \frac{1}{\epsilon_{i,j}^2}\mathds{I}_D \right)^{-1}$ \\
$ \bmu_{i,j}^{\gamma} = c_{i,j}^{2} \bLam^{\gamma}_{i,j} \bsQ^{\top} \bSig^{-1}  \left(\bsy_{i,j}- c_{i,j}\bsp_{i,j}\right)$ \\
$\bLam_{r,i,j}^{k} = \left( \frac{1}{\beta_r^2 } \mathds{I}_L
+ \frac{1}{\alpha_r^2} \bsH^{-1} + a_{r,i,j}^2 \bSig^{-1} \right)^{-1}$ \\
$\bmu_{r,i,j}^{k} = \bLam_{r,i,j}^{k} \left(a_{r,i,j}^2 \bSig^{-1} \omega_{r,i,j} + \frac{1}{\beta_r^2 } \bmu_{r,i,j}\right)$  \\
$\bLam_{i,j}^{d} = \left(\bSig^{-1} + \frac{1}{\epsilon_{i,j}^2} \bsH^{-1}  \right)^{-1} $ \\
$\bmu_{i,j}^{d} = \bLam_{i,j}^{d}\bSig^{-1} \left(\bsy_{i,j} - c_{i,j} \bsp_{i,j}\right)$  \\ 
$\bLam_{i,j}^{c} = \left(\frac{1}{\eta^2} + \bsp_{i,j}^{\top}\bsp_{i,j} \right)^{-1} $ \\
$\bmu_{i,j}^{c} = \bLam_{i,j}^{c}  \left[\bsp_{i,j}^{\top}  \bSig^{-1} \left(\bsy_{i,j} - \bphi_{i,j}\right) + \frac{1}{\eta^2} \right]$
\end{tabular}\right.
\end{equation}
with   $\bsQ =    \left(\bsm_{1}\odot\bsm_{1}, \cdots,\bsm_{R}\odot\bsm_{R},\sqrt{2} \bsm_{1}\odot\bsm_{2}, \cdots, \right.$ $\left. \sqrt{2} \bsm_{R-1}\odot\bsm_{R}\right)$, $\bsp_{i,j} = \bsM \bsa_{i,j}$,  
$\omega_{r,i,j} = \frac{1}{a_{r,i,j}} \left(\bsy_{i,j} - \bsp_{i,j} -\sum_{r'\neq r}{ a_{r',i,j} \bsk_{r',i,j}}\right)$, $\bsz_{i,j} = \bsQ_{i,j} \bgam_{i,j}$, $x_0 = \bsy_{i,j}^{\top} \bSig^{-1} \bsy_{i,j}$, $x_1 = \bsy_{i,j}^{\top} \bSig^{-1} \bsp_{i,j}$, $x_2 =0.5 \left( \bsp_{i,j}^{\top} \bSig^{-1} \bsp_{i,j} - 2 \bsz_{i,j}^{\top} \bSig^{-1} \bsy_{i,j} \right)$, $x_3 = \bsz_{i,j}^{\top} \bSig^{-1} \bsp_{i,j}$ and $x_4 =0.5 \bsz_{i,j}^{\top} \bSig^{-1} \bsz_{i,j}$. Note that $ f\left(w_{i,j}^2 | \bThe_{\backslash w_{i,j}^2} \right)$ is given by \eqref{eqt:CondGam_IGam}. 
The independence between the parameters leads to $f\left(\bsA | \bThe_{\backslash \bsA} \right) = \prod_{i,j} f\left(\bsa_{i,j} | \bThe_{\backslash \bsa_{i,j}} \right)$, $f\left(\bGam | \bThe_{\backslash \bGam} \right) = \prod_{i,j} f\left(\bgam_{i,j} | \bThe_{\backslash \bgam_{i,j}} \right)$, $f\left(\bsD | \bThe_{\backslash \bsD} \right) =\prod_{i,j} f\left(\bsd_{i,j} | \bThe_{\backslash \bsd_{i,j}} \right)$, $f\left(\bsc | \bThe_{\backslash \bsc} \right) = \prod_{i,j} f\left(c_{i,j} | \bThe_{\backslash c_{i,j}} \right)$, $f\left(\bSig | \bThe_{\backslash \bSig} \right) = \prod_{\ell} f\left(\sigma^2_{\ell} | \bThe_{\backslash \sigma^2_{\ell}} \right)$, $f\left(\beps| \bThe_{\backslash \beps} \right) = \prod_{i,j} f\left(\epsilon^2_{i,j} | \bThe_{\backslash \epsilon^2_{i,j}} \right)$, and $f\left(\bsw | \bThe_{\backslash \bsw} \right) = \prod_{i,j} f\left(w_{i,j}^2 | \bThe_{\backslash w_{i,j}^2} \right)$. Note finally that the matrix $\bsK$ is updated iteratively using a checkerboard scheme.

\bibliographystyle{ieeetran}
\bibliography{biblio_all}

\end{document}